# Statistical Issues on the Neutrino Mass Hierarchy with $\Delta\chi^2$

F. Sawy[1] and L. Stanco[2]

[1] American University in Cairo, Cairo, Egypt
[2] INFN, Sezione di Padova, 35131 Padova, Italy

September 4, 2023

## Abstract

The Neutrino Mass Hierarchy Determination ($\nu$ MHD) is one of the main goals of the major current and future neutrino experiments. The statistical analysis usually proceeds from a standard method, a single dimensional estimator ($1D - \Delta\chi^2$) that shows some draw-backs and concerns, together with a debatable strategy. The draw-backs and considerations of the standard method will be explianed through the following three main issues. First issue is the limited power of the standard method. The $\Delta\chi^2$ estimator provides us with different results when different simulation procedures were used. Second issue, when $\chi^2_{min(NH)}$ and $\chi^2_{min(IH)}$ are drawn in a $2D$ map, their strong positive correlation manifests $\chi^2$ as a bi-dimensional instead of single dimensional estimator. The overlapping between the $\chi^2$ distributions of the two hypotheses leads to the experiment sensitivity reduction. Third issue is the robustness of the standard method. When the JUNO sensitivity is obtained using different procedures, $\Delta\chi^2$ as one dimensional and $\chi^2$ as two dimensional estimator, the experimental sensitivity varies with the different values of the atmospheric mass, the input parameter. We computed the oscillation of $|\overline{\Delta\chi^2}|$ with the input parameter values, $|\Delta m^2|_{input}$. The MH significance using the standard method, $\Delta\chi^2$, strongly depends on the values of the parameter $|\Delta m^2|_{input}$. Consequently, the experiment sensitivity depends on the precision of the atmospheric mass. This evaluation of the standard method confirms the draw-backs.

## 1 Introduction

Neutrino oscillations is a quantum mechanical phenomenon in which neutrino flavor changes spontaneously to another flavor. According to the standard 3 neutrino paradigm, neutrinos come with three flavors, $\nu_e$, $\nu_\mu$ and $\nu_\tau$, and with three $\nu_1$, $\nu_2$ and $\nu_3$ mass eigenstates [1]. Although neutrinos were introduced over 80 years ago, their properties remain to a large extent unknown [2]. Some of the $3\nu$ paradigm fundamental parameters are still missing until now like: the absolute masses of neutrinos[3], the amount of the possible leptonic charge parity violation (CPV)[4], Dirac or Majorana neutrino nature [5] and the neutrino mass ordering [6].

Currently, the determination of the neutrino mass ordering using reactor neutrino spectrum is pursued by several experiments and proposals. There are some challenges facing anyone that tries to solve this problem. First, its evaluation from reactor experiments is based on the tiny interference effect between the $\Delta m^2_{31}$ and $\Delta m^2_{23}$ oscillations [7]. Second, current analyses require several years of data taking and an extreme energy resolution to achieve anyhow less than $5\,\sigma$. Third, the sensitivity may depend on the input values of the oscillation parameters used by the global fits on the oscillation analysis. In particular, the neutrino atmospheric mass may have different values for normal ordering (NH) or inverted one (IH). The answer to the third point depends on the used analysis method. It is mandatory to establish the robustness of all these analyses.

The Jiangmen Underground Neutrino Observatory (JUNO) experiment [8] has been proposed and approved for realization in the south of China, being the mass ordering evaluation one of its main goals. JUNO will allow to single out one of the missing fundamental information, the neutrino mass hierarchy, in an almost independent way of the other neutrino parameters. In particular, there will be no dependence on the phase of the leptonic CP violation, $\delta_{CP}$, no strong dependence on three vs four neutrino pattern, no dependence on $\theta_{13}$, no dependence on matter effects [8]. The mass hierarchy study can be performed by looking at the vacuum oscillation pattern in medium baselines reactor anti-neutrino experiments [9]. The JUNO strategy is based on the observation that the contribution to the oscillation probability is represented by fast oscillating terms superimposed to a general oscillation pattern. Their relative size changes according to the two different possibilities, NH or IH, leading to a contribution of opposite sign in the two cases. Therefore, it is possible to discriminate between the two possible mass hierarchies by studying the interference between the two oscillation frequencies driven by $\Delta m_{31}^2$ and $\Delta m_{23}^2$ in the reactor antineutrino spectrum [10]. The discrimination power of the experiment is maximized when the $\Delta m_{21}^2$ oscillation is maximal, and the baseline at JUNO has been chosen in such a way to realize this condition [11]. Since the difference of neutrino oscillation in vacuum for different mass hierarchies is very small, energy resolution is the crucial factor for the success of JUNO. The goal is that the energy resolution reaches $3\%/\sqrt{E}$ at 1 MeV to detect electron neutrino coming from reactor plants.

In the next section, the so usual $\chi^2$ method is recollected and evaluated. In the following sections, the three issues of the standard algorthiem are explianed. Section three includes the first issue, section four includes the second issue then third issue is included in the fifth section. Then the results is demonstrated in the sixth section. After that the conclusions are indicated in seventh section. Finally, in the Appendix a technical description of the implementation of the simulation is reported.

## 2 The Standard Method

A different definition with of the $\chi^2$ function based on the Poisson distribution yields a consistent MH sensitivity [8]. For JUNO, $\chi^2$ is divided into three parts as indicated in

$$\chi^2 = \chi_{para}^2 + \chi_{sys}^2 + \chi_{stat}^2. \tag{1}$$

$\chi_{para}^2$ summarizes the prior knowledge on oscillation parameters. In JUNO these parameters are $\sin^2 2\theta_{12}, \sin^2 2\theta_{13}, \Delta m_{12}^2$ and $\Delta m_{13}^2$. Then $\chi_{para}^2$ becomes:

$$\begin{aligned}\chi_{para}^2 = & \left(\frac{(\sin^2 2\theta_{12})^{fit} - (\sin^2 2\theta_{12})^{input}}{\sigma_{\sin^2 2\theta_{12}}}\right)^2 \\ & + \left(\frac{(\sin^2 2\theta_{13})^{fit} - (\sin^2 2\theta_{13})^{input}}{\sigma_{\sin^2 2\theta_{13}}}\right)^2 \\ & + \left(\frac{(|\Delta m_{31}^2|)^{fit} - (|\Delta m_{31}^2|)^{input}}{\sigma_{|\Delta m_{31}^2|}}\right)^2 \\ & + \left(\frac{(\Delta m_{21}^2)^{fit} - (\Delta m_{21}^2)^{input}}{\sigma_{\Delta m_{21}^2}}\right)^2.\end{aligned} \tag{2}$$

The reactor anti-neutrino flux, the inverted beta decay cross section, the fiducial volume and the weight fraction of free proton can all be combined into a single overall factor. Consequently, their contributions to the $\chi^2$ function can be represented by a single term as,

$$\chi^2_{sys} = \left( \frac{f^{fit}_{sys} - f^{input}_{sys}}{\sigma_{f_{sys}}} \right)^2, \quad (3)$$

where $f^{input}_{sys} = 1$, and $\sigma_{f_{sys}} = 0.03$.

The last term of Equation 1, $\chi^2_{stat}$, represents the statistical fluctuation. When we introduce binning with respect to $E^{obs}_{vis}$, it looks like

$$\chi^2_{stat} = \sum_i \left( \frac{N^{fit}_i - N^{NH(IH)}_i}{\sqrt{N^{NH(IH)}_i}} \right)^2 \quad (4)$$

with the summation running over all the bins. Here, $N^{NH(IH)}_i$ is the event number for the $i_{th}$ bin when the hierarchy is NH(IH). $N^{fit}_i$ is the fitted number of events, calculated as a function of the four model parameters and the normalization factor $f_{sys}$. All parameters are varied under the NH(IH) constraints of Equation 2 and Equation 3.

In the minimization procedure all the parameters were initially set to their global best values that are indicated in Table 1. The fitting procedures and the minimization of $\chi^2$ are done with the TMinuit algorithm (ROOT libraries). The $\chi^2$ distributions are obtained for four parameters ($sin^2\theta_{12}$, $sin^2\theta_{13}$, $\delta m^2_{sol}$ and $\Delta m^2$), based on a total of 108357 signal events (Figure 1 and Figure 2).

**Table 1:** *The recent best-fit values for the oscillation parameters, as indicated in [12].*

|  | best-fit | $3\sigma$ region |
|---|---|---|
| $Sin^2_{12}$ | 0.2970 | 0.2500 - 0.3540 |
| $Sin^2_{13}(NH)$ | 0.02140 | 0.0185 - 0.0246 |
| $Sin^2_{13}(IH)$ | 0.02180 | 0.0186 - 0.0248 |
| $\delta m^2_{sol}$ | $7.37 \times 10^{-5}$ | $6.93 \times 10^{-5} - 7.97 \times 10^{-5}$ |
| $\Delta m^2(NH)$ | $2.500 \times 10^{-3}$ | $2.37 \times 10^{-3}$ - $2.63 \times 10^{-3}$ |
| $\Delta m^2(IH)$ | $2.460 \times 10^{-3}$ | $-2.60 \times 10^{-3}$ to $-2.33 \times 10^{-3}$ |

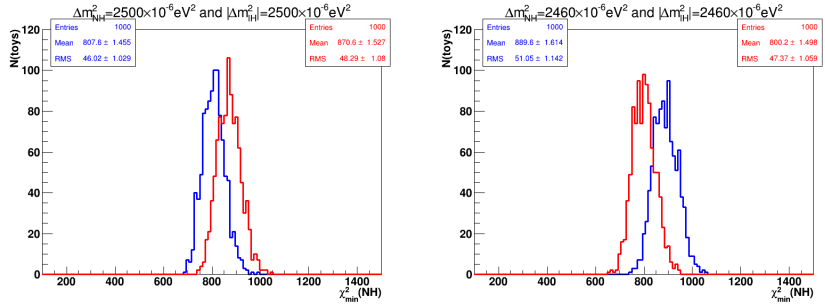

**Figure 1:** *Two $\chi^2$ distributions for 1000 (NH) + 1000 (IH) toy JUNO-like simulations generated at $\Delta m^2 = 2.500 \times 10^{-3} eV^2$ for NH hypothesis (left panel) and $\Delta m^2_{input} = -2.460 \times 10^{-3} eV^2$ for IH hypothesis (right panel), with six years of exposure and the ten near reactor cores with infinite energy resolution. The intrinsic strong positive correlation between the two components $\chi^2_{min(NH)}$ and $\chi^2_{min(IH)}$ leads to the overlapping between the two $\chi^2$ distributions.*

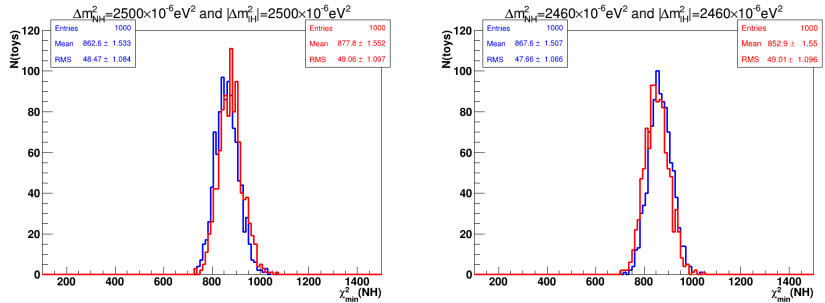

**Figure 2:** *Two $\chi^2$ distributions for 1000 (NH) + 1000 (IH) toy JUNO-like simulations generated at $\Delta m^2_{input} = 2.500 \times 10^{-3}\text{eV}^2$ for NH hypothesis (left plot) and $\Delta m^2_{input} = -2.460 \times 10^{-3}\text{eV}^2$ for IH hypothesis (right panel) with six years of exposure and the ten near reactor cores, with 3% relative energy resolution. The intrinsic strong positive correlation between the two components $\chi^2_{min(NH)}$ and $\chi^2_{min(IH)}$ leads to an overlapping between the two $\chi^2$ distributions.*

As reported in [8], the sensitivity can reach $|\overline{\Delta\chi^2}| > 16$ in the ideal case of single reactor and single detector, and $|\overline{\Delta\chi^2}| > 9$ considering the spread of reactor cores and uncertainties of the detector response. All these results has been reached using semi-analytical simulations i.e. simulations as used in [8] and [13]. Semi-Analytical simulations are generated by fluctuating the bin-content according to Poisson or Gaussian distributions that represent the number of events. In addition, a second fluctuation is added by applying $3\%/\sqrt{E}$ energy smearing in each single bin not in each single event. If the energy resolution smearing per each single event is replaced by smearing for the whole bin, an event balance migration occurs and the number of events per each single bin becomes uncorrelated with side bins leading to the results reported in the [8]. We provided the simulation performed on a event-by-event basis and computed the experimental sensitivity for the JUNO by changing the atmospheric neutrino mass. The $\chi^2$ distributions are provided for $\Delta m^2_{input} = -2.460 \times 10^{-3}\text{eV}^2$ and $\Delta m^2_{input} = 2.500 \times 10^{-3}\text{eV}^2$ for IH hypothesis and NH hypothesis respectively(Figure 1 and Figure 2) with infinite and 3% relative energy resolution respectively.

## 3 Issue one: the Limited Power of $\Delta\chi^2$ as a Single Dimensional Estimator

The two discrete hypotheses are not nested, thus the Wilks theorem is not applicable in this problem when it is based on the $\Delta\chi^2$ defined in Equation 5. As a consequence, $\Delta\chi^2$ does not follow a $\chi^2$ distribution [14]. The MO significance is usually obtained in terms of the single dimensional estimator $\Delta\chi^2$ and its evaluation is based on two distinct hypotheses, NH and IH. For each MO the best solution is found: the $\chi^2_{min}$ comes from two different best-fit values for NH model to be $\chi^2_{min(NH)}$ and IH model to be $\chi^2_{min(IH)}$:

$$\Delta\chi^2 = \chi^2_{min(NH)} - \chi^2_{min(IH)}, \qquad (5)$$

where the two minima are evaluated spanning the uncertainties on the three-neutrino oscillation parameters. The experimental sensitivity to the neutrino mass hierarchy arises from the small phase shift in the oscillation terms depending on the two large mass-squared differences $\Delta m^2_{32}$ and $\Delta m^2_{31}$. JUNO sensitivity can be calculated using the single dimensional test statistics $\Delta\chi^2$. The median sensitivity can be obtained using the Z-test, where $z^{(NH)}_{score}$ is the number of $\sigma_{NH}$ assuming NH be the true model and $z^{(IH)}_{score}$ is the number of $\sigma_{IH}$ assuming IH be the true model,

$$z_{score}^{(NH)} = \frac{\overline{\Delta\chi^2}^{(IH)} - \overline{\Delta\chi^2}^{(NH)}}{\sigma_{NH}} \qquad z_{score}^{(IH)} = \frac{\overline{\Delta\chi^2}^{(NH)} - \overline{\Delta\chi^2}^{(IH)}}{\sigma_{IH}}. \qquad (6)$$

The $\overline{\Delta\chi^2}^{(NH)}$, $\sigma_{NH}$, $\overline{\Delta\chi^2}^{(IH)}$ and $\sigma_{IH}$ are the mean value and standard deviation of the $\Delta\chi^2$ distribution assuming NH and IH be the true model, respectively. There an approximation is usually used [8, 15, 16, 17]:

$$\sigma_{\Delta\chi^2} = 2\sqrt{\overline{\Delta\chi^2}}, \qquad (7)$$

therefore, Equation 6 becomes:

$$z_{score}^{(NH)} = \sqrt{\overline{\Delta\chi^2}^{(NH)}} \qquad z_{score}^{(IH)} = \sqrt{\overline{\Delta\chi^2}^{(IH)}}. \qquad (8)$$

When the analysis is performed on a event-by-event basis and not a semi-analytical simulations as in [8], the dispersions of the distributions cannot be described by Equation 7 anymore. That significantly affect the statistical significance that drops to less than $2\,\sigma$ as indicated in Table 3. The reason stays in the convolution of the energy resolution. To chick it: the analysis is done also at an infinite energy resolution to find out whether it works (Figure 3).

The investigation of the origin of the approximation has been pursued by looking whether it is still valid in event-by-event simulations as it is in semi-analytical simulations. In fact, we found that the dispersion of the two distributions becomes wider than in semi-analytical simulations when an finite energy resolution is taken into account. The energy error introduces strong correlations between bins and it corresponds to a further systematic error.

The limited power of the $\Delta\chi^2$ manifests itself being controlled by the statistical assumption i.e. Equation 7. The experimental sensitivity is reduced when the energy systematic error is taken into account, and Equation 7 is no more valid. Specific cases are reported in the following figures and tables and other details are reported in subsection 6.1.

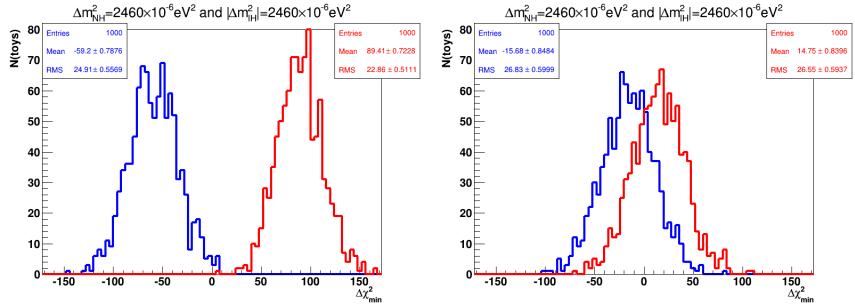

**Figure 3:** $\Delta\chi^2$ estimator for 1000 (NH) + 1000 (IH) toy JUNO-like simulations generated at $|\Delta m^2| = 2.460 \times 10^{-3}$eV$^2$ for NH and IH hypotheses with six years of exposure and the ten near reactor cores. An infinite energy resolution is assumed for the left plot and 3% relative energy resolution for the right plot. The experimental sensitivities under these terms are reported in Table 3 and Table 2.

**Table 2:** *The comparison of the MH sensitivity at infinite energy resolution for NH sample and IH sample at $|\Delta m^2| = 2.460 \times 10^{-3} \text{eV}^2$ in two cases. The first case uses Equation 6 and the second one uses Equation 8.*

| Infinite Energy Resolution | | |
|---|---|---|
| $\mu_{NH}$ | $-59.20 \pm 0.79$ | |
| $\sigma_{NH}$ | $24.91 \pm 0.56$ | |
| $\mu_{IH}$ | $89.41 \pm 0.72$ | |
| $\sigma_{IH}$ | $22.86 \pm 0.51$ | |
| $z_{score}^{(NH)}$ | 5.966 | 7.694(app.) |
| $z_{score}^{(IH)}$ | 6.501 | 9.456(app.) |

**Table 3:** *The comparison of the MH sensitivity at energy resolution $3\%/\sqrt{E}$ for NH sample and IH sample at $|\Delta m^2| = 2.460 \times 10^{-3} \text{eV}^2$ in two cases. The first case uses Equation 6 and the second one uses Equation 8.*

| energy resolution $3\%/\sqrt{E}$ | | |
|---|---|---|
| $\mu_{NH}$ | $-15.68 \pm 0.85$ | |
| $\sigma_{NH}$ | $26.83 \pm 0.60$ | |
| $\mu_{IH}$ | $14.75 \pm 0.84$ | |
| $\sigma_{IH}$ | $26.55 \pm 0.60$ | |
| $z_{score}^{(NH)}$ | 1.134 | 3.960(app.) |
| $z_{score}^{(IH)}$ | 1.146 | 3.841(app.) |

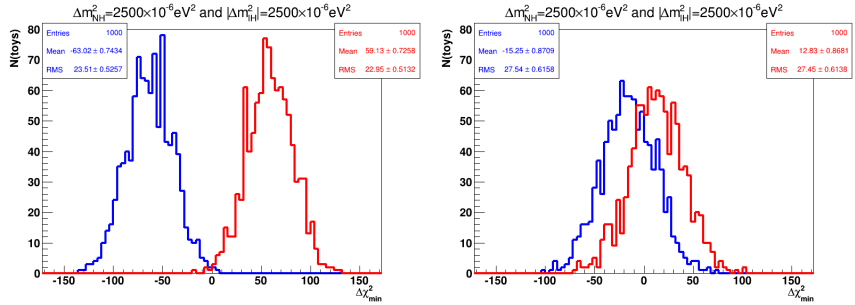

**Figure 4:** *$\Delta\chi^2$ estimator for 1000 (NH) + 1000 (IH) toy JUNO-like simulations generated at $|\Delta m^2| = 2.500 \times 10^{-3} \text{eV}^2$ for NH hypothesis (left panel) and IH hypothesis (right panel) with six years of exposure and the ten near reactor cores. An infinite energy resolution is assumed for the left plot and 3% relative energy resolution for the right plot. The experimental sensitivities under these terms are reported in Table 4 and Table 5, respectively.*

**Table 4:** *The comparison of the MH sensitivity for ideal distributions for NH sample and IH sample at $|\Delta m^2| = 2.500 \times 10^{-3} \text{eV}^2$ in two cases. The first case uses Equation 6 and the second one uses Equation 8.*

| | Infinite Energy Resolution | |
|---|---|---|
| $\mu_{NH}$ | $-63.02 \pm 0.74$ | |
| $\sigma_{NH}$ | $23.51 \pm 0.53$ | |
| $\mu_{IH}$ | $59.13 \pm 0.73$ | |
| $\sigma_{IH}$ | $22.95 \pm 0.51$ | |
| $z_{score}^{(NH)}$ | 5.203 | 7.950(app.) |
| $z_{score}^{(IH)}$ | 5.330 | 7.690(app.) |

**Table 5:** *The comparison of the MH sensitivity for actual distributions for NH sample and IH sample at $|\Delta m^2| = 2.500 \times 10^{-3} \text{eV}^2$ in two cases. The first case uses Equation 6 and the second one uses Equation 8.*

| | $3\%/\sqrt{E}$ Energy Resolution | |
|---|---|---|
| $\mu_{NH}$ | $-15.25 \pm 0.87$ | |
| $\sigma_{NH}$ | $27.54 \pm 0.62$ | |
| $\mu_{IH}$ | $12.83 \pm 0.87$ | |
| $\sigma_{IH}$ | $27.45 \pm 0.61$ | |
| $z_{score}^{(NH)}$ | 1.020 | 3.901(app.) |
| $z_{score}^{(IH)}$ | 1.023 | 3.582(app.) |

Figure 5 for NH sample at $\Delta m^2 = 2.500 \times 10^{-3} \text{eV}^2$ and IH sample for $\Delta m^2 = -2.460 \times 10^{-3} \text{eV}^2$, shows the $\Delta \chi^2$ distributions for a relative 3% and an infinite energy resolution. The JUNO sensitivity is clearly different from that reported in [8].

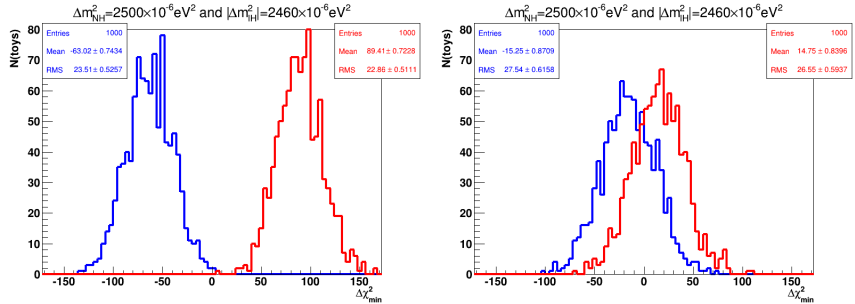

**Figure 5:** *$\Delta \chi^2$ estimator for 1000 (NH) + 1000 (IH) toy JUNO-like simulations generated at $\Delta m^2 = 2.500 \times 10^{-3} \text{eV}^2$ for NH hypothesis (blue) and $\Delta m^2 = -2.460 \times 10^{-3} \text{eV}^2$ for IH hypothesis (red) with six years of exposure and the ten near reactor cores. The left plot is for infinite energy resolution and the blue plot is for 3% relative energy resolution. The experimental sensitivities under these terms are reported in Table 6 and Table 7.*

**Table 6:** *The comparison of the MH sensitivity for ideal distributions for NH sample at $\Delta m^2 = 2.500 \times 10^{-3} \text{eV}^2$ and IH sample for $\Delta m^2 = -2.460 \times 10^{-3} \text{eV}^2$ in two cases. The first case uses Equation 6 and the second one uses Equation 8.*

| Infinite Energy Resolution | | |
|---|---|---|
| $\mu_{NH}$ | $-63.02 \pm 0.74$ | |
| $\sigma_{NH}$ | $23.51 \pm 0.53$ | |
| $\mu_{IH}$ | $89.41 \pm 0.72$ | |
| $\sigma_{IH}$ | $22.86 \pm 0.51$ | |
| $z_{score}^{(NH)}$ | 6.484 | 7.950(app.) |
| $z_{score}^{(IH)}$ | 6.668 | 9.456(app.) |

When only statistical fluctuations are included, the MH sensitivities using Z-test ($z_{score}^{(NH)}$ and $z_{score}^{(IH)}$) do not exact equal to the MH sensitivities obtained in the approximated Equation 7($z_{score}^{(NH)}(app.)$ and $z_{score}^{(IH)}(app.)$) as reported in Table 4. This observation is consistent with what obtained at the atmospheric mass, $|\Delta m^2| = 2.460 \times 10^{-3} \text{eV}^2$ as reported in Table 6. This conclusion will be confirmed for other 18 different values for the atmospheric mass at infinite energy resolution in subsection 6.1.

**Table 7:** *The comparison of the MH sensitivity for actual distributions for NH sample at $\Delta m^2 = 2.500 \times 10^{-3} \text{eV}^2$ and IH sample for $\Delta m^2 = -2.460 \times 10^{-3} \text{eV}^2$ in two cases. The first case uses Equation 6 and the second one uses Equation 8.*

| $3\%/\sqrt{E}$ energy resolution | | |
|---|---|---|
| $\mu_{NH}$ | $-15.25 \pm 0.87$ | |
| $\sigma_{NH}$ | $27.54 \pm 0.62$ | |
| $\mu_{IH}$ | $14.75 \pm 0.84$ | |
| $\sigma_{IH}$ | $26.55 \pm 0.60$ | |
| $z_{score}^{(NH)}$ | 1.089 | 3.960(app.) |
| $z_{score}^{(IH)}$ | 1.130 | 3.841(app.) |

## 4 Issue two: non-bright Results Using $\chi^2$ as a Bi-Dimensional Estimator

When $\chi^2_{min(IH)}$ and $\chi^2_{min(NH)}$ are drawn in $2D$ dimensional map, their strong positive correlation manifests $\chi^2$ as a bi-dimensional estimator. This strong positive correlation leads to overlap between the $\chi^2$ distributions of the two hypotheses, thus reducing the experiment sensitivity. When we look at $\chi^2$ as a bi-dimensional estimator, the experiment sensitivity can be calculated with a Z-test for two dimensional test statistic providing the results indicated in Table 8 and Table 9.

Using Z-test for $2D$, the MH sensitivity can be calculated as

$$z_{score}^{(NH)} = \frac{\sqrt{(\overline{\chi^2}_{IH}^{(NH)} - \overline{\chi^2}_{IH}^{(IH)})^2 + (\overline{\chi^2}_{NH}^{(NH)} - \overline{\chi^2}_{NH}^{(IH)})^2}}{\sqrt{(\sigma_{IH}^2)^{NH} + (\sigma_{NH}^2)^{NH}}}$$

$$z_{score}^{(IH)} = \frac{\sqrt{(\overline{\chi^2}_{IH}^{(IH)} - \overline{\chi^2}_{IH}^{(NH)})^2 + (\overline{\chi^2}_{NH}^{(IH)} - \overline{\chi^2}_{NH}^{(NH)})^2}}{\sqrt{(\sigma_{IH}^2)^{IH} + (\sigma_{NH}^2)^{IH}}}$$

(9)

where $\overline{\chi^2}_{IH}^{(NH)}$ and $(\sigma_{IH}^2)^{NH}$ are the mean and the standard derivation of $\chi^2$ distribution of the NH sample assuming IH hypothesis is the true hypothesis. $\overline{\chi^2}_{NH}^{(NH)}$ and $(\sigma_{NH}^2)^{NH}$ are the mean and the standard derivation of $\chi^2$ distribution of the NH sample assuming NH hypothesis is the true hypothesis. $\overline{\chi^2}_{IH}^{(IH)}$ and $(\sigma_{IH}^2)^{IH}$ are the mean and the standard deviation of $\chi^2$ distribution of the IH sample assuming IH hypothesis is the true hypothesis. $\overline{\chi^2}_{NH}^{(IH)}$ and $(\sigma_{NH}^2)^{IH}$ are the mean and the standard deviation of $\chi^2$ distribution of the IH sample assuming NH hypothesis is the true hypothesis.

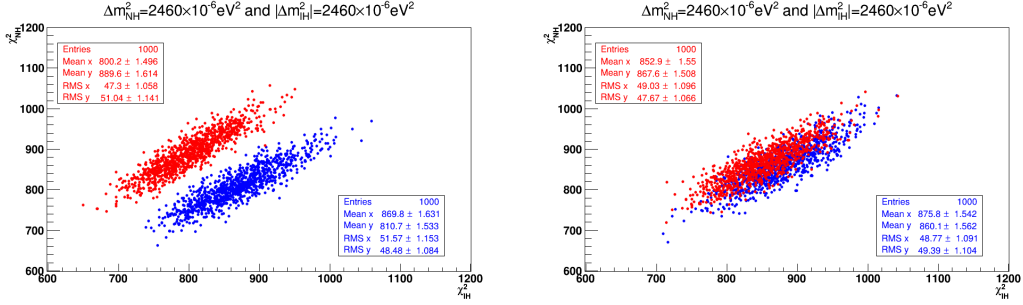

**Figure 6:** *Two islands of $\chi^2$ for 1000 (NH) + 1000 (IH) toy JUNO-like simulations generated at $|\Delta m^2| = 2.460 \times 10^{-3} \text{eV}^2$ for NH hypothesis (blue color) and IH hypothesis (red color) with six years of exposure and the ten near reactor cores. An infinite energy resolution is assumed for left plot and $3\%/\sqrt{E}$ energy resolution for right plot. The experimental sensitivities under these terms are reported in Table 8.*

**Table 8:** *Two $\chi^2$ distributions for 1000 (NH) + 1000 (IH) toy JUNO-like simulations generated at $|\Delta m^2| = 2.460 \times 10^{-3} \text{eV}^2$ for NH and IH hypotheses with six years of exposure and the ten near reactor cores. The sensitivity is calculated using Equation 9.*

| Energy resolution | infinite | | 3% | |
|---|---|---|---|---|
| | **NH** | **IH** | **NH** | **IH** |
| $\mu_{NH}$ | $810.7 \pm 1.53$ | $889.6 \pm 1.61$ | $860.10 \pm 1.56$ | $867.60 \pm 1.51$ |
| $\sigma_{NH}$ | $48.48 \pm 1.08$ | $51.05 \pm 1.14$ | $49.39 \pm 1.10$ | $47.67 \pm 1.06$ |
| $\mu_{IH}$ | $869.8 \pm 1.63$ | $800.2 \pm 1.50$ | $875.80 \pm 1.54$ | $852.9 \pm 1.55$ |
| $\sigma_{IH}$ | $51.57 \pm 1.15$ | $47.30 \pm 1.06$ | $48.77 \pm 1.09$ | $49.03 \pm 1.10$ |
| $z_{score}^{(NH)}$ | $1.072\,\sigma$ | | $0.219\,\sigma$ | |
| $z_{score}^{(IH)}$ | $1.089\,\sigma$ | | $0.223\,\sigma$ | |

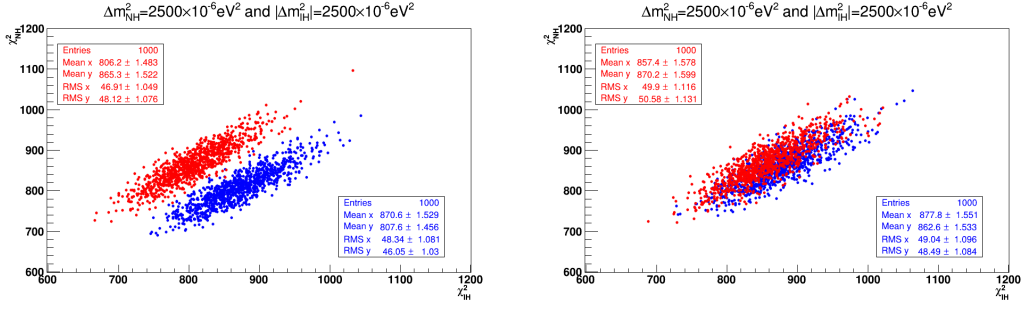

**Figure 7:** *Two islands of $\chi^2$ for 1000 (NH) + 1000 (IH) toy JUNO-like simulations generated at $|\Delta m^2| = 2.500 \times 10^{-3} \text{eV}^2$ for NH hypothesis (blue island) and IH hypothesis (red island) with six years of exposure and the ten near reactor cores. An infinite energy resolution is assumed for left plot and 3% relative energy resolution for right plot. The experimental sensitivities under these terms are reported in Table 9.*

**Table 9:** *Two $\chi^2$ distributions for 1000 (NH) + 1000 (IH) toy JUNO-like simulations generated at $|\Delta m^2| = 2.500 \times 10^{-3} \text{eV}^2$ for NH and IH hypotheses with six years of exposure and ten near reactor cores. The sensitivity is calculated using Equation 9.*

|  | infinite | | 3% | |
|---|---|---|---|---|
|  | **NH** | **IH** | **NH** | **IH** |
| $\mu_{NH}$ | $807.6 \pm 1.46$ | $865.30 \pm 1.52$ | $862.60 \pm 1.53$ | $870.20 \pm 1.60$ |
| $\sigma_{NH}$ | $46.05 \pm 1.03$ | $48.12 \pm 1.08$ | $48.49 \pm 1.08$ | $50.58 \pm 1.13$ |
| $\mu_{IH}$ | $870.60 \pm 1.53$ | $806.20 \pm 1.48$ | $877.80 \pm 1.55$ | $857.4 \pm 1.58$ |
| $\sigma_{IH}$ | $48.34 \pm 1.08$ | $46.91 \pm 1.05$ | $49.04 \pm 1.10$ | $49.90 \pm 1.12$ |
| $z_{score}^{(NH)}$ | $0.916\,\sigma$ | | $0.204\,\sigma$ | |
| $z_{score}^{(IH)}$ | $0.910\,\sigma$ | | $0.200\,\sigma$ | |

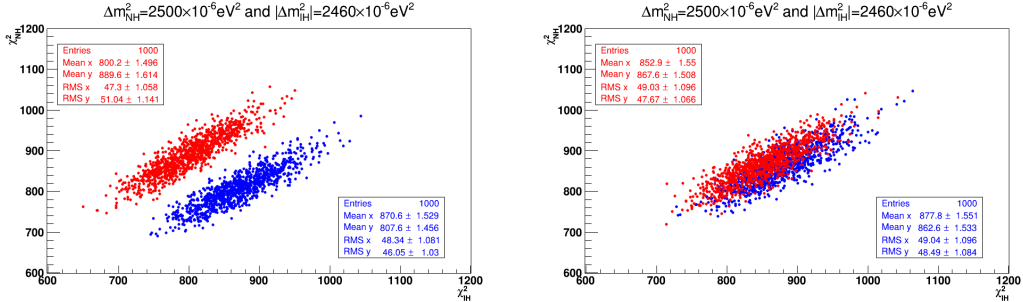

**Figure 8:** *Two islands of $\chi^2$ for 1000 (NH) + 1000 (IH) toy JUNO-like simulations generated at $\Delta m^2 = 2.500 \times 10^{-3} \text{eV}^2$ for NH hypothesis (blue island) and $\Delta m^2 = -2.460 \times 10^{-3} \text{eV}^2$ for IH hypothesis (red island) with six years of exposure and the ten near reactor cores. An infinite energy resolution is assumed for left plot and 3% relative energy resolution for right plot. The experimental sensitivities under these terms are reported in Table 10.*

**Table 10:** *Two $\chi^2$ distributions for 1000 (NH) + 1000 (IH) toy JUNO-like simulations generated at $\Delta m^2 = 2.500 \times 10^{-3} \text{eV}^2$ for NH hypothesis and $\Delta m^2 = -2.460 \times 10^{-3} \text{eV}^2$ for IH hypothesis with six years of exposure and ten near reactor cores. The sensitivity is calculated using Equation 9.*

|  | infinite | | 3% | |
|---|---|---|---|---|
|  | **NH** | **IH** | **NH** | **IH** |
| $\mu_{NH}$ | $807.6 \pm 1.46$ | $889.6 \pm 1.61$ | $862.60 \pm 1.53$ | $867.6 \pm 1.51$ |
| $\sigma_{NH}$ | $46.05 \pm 1.03$ | $51.05 \pm 1.14$ | $48.49 \pm 1.08$ | $47.67 \pm 1.07$ |
| $\mu_{IH}$ | $870.60 \pm 1.53$ | $800.2 \pm 1.50$ | $877.80 \pm 1.55$ | $852.90 \pm 1.55$ |
| $\sigma_{IH}$ | $48.34 \pm 1.08$ | $47.30 \pm 1.06$ | $49.04 \pm 1.08$ | $49.03 \pm 1.07$ |
| $z_{score}^{(NH)}$ | $1.159\,\sigma$ | | $0.217\,\sigma$ | |
| $z_{score}^{(IH)}$ | $1.113\,\sigma$ | | $0.219\,\sigma$ | |

# 5 Issue Three: the robustness

Robust statistics are the statistics that yield good performance when data is drawn from a wide range of probability distributions that are largely unaffected by outliers or small departures from model assumptions in a given data set [18]. In other words, a robust statistic is resistant to initial deviations with respect to the final results [19].

The main focus of the statistical analysis using the $\Delta\chi^2$ standard method is to calculate neutrino mass hierarchy determination sensitivity and less attention or none is put about its robustness. subsection 5.1 will discuss how the standard method using $\Delta\chi^2$ is not able to maintain the robustness while subsection 5.2 will discuss inability of the $\chi^2$ to establish the robustness as a bi-dimensional estimator. This study is done for 20 different data values of the input atmospheric neutrino mass in the range, $2.450 \times 10^{-3}\text{eV}^2 \leq |\Delta m^2|_{input} \leq 2.580 \times 10^{-3}\text{eV}^2$.

## 5.1 The $|\overline{\Delta\chi^2}|$ oscillations with $\Delta m^2_{input}$

There are trends in our data to confirm that the $|\overline{\Delta\chi^2}|$ varies with the input atmospheric neutrino mass $|\Delta m^2|_{input}$. We studied the relation between the $|\overline{\Delta\chi^2}|$ values and the value of the input parameter for 20 different values, $|\Delta m^2|_{input}$ in the range, $2.450 \times 10^{-3}\text{eV}^2 \leq |\Delta m^2|_{input} \leq 2.580 \times 10^{-3}\text{eV}^2$ and we computed the corresponding experimental sensitivity for the two cases with/without including the systematic uncertainties. Figure 9 illustrates the variation of $|\overline{\Delta\chi^2}|$ as a function of the input atmospheric neutrino mass $|\Delta m^2|_{input}$, in the range of $2.450 \times 10^{-3}\text{eV}^2 \leq |\Delta m^2|_{input} \leq 2.580 \times 10^{-3}\text{eV}^2$, assuming infinite energy resolution. Figure 10 illustrates the variation of $\Delta\chi^2$ with the input atmospheric neutrino mass $|\Delta m^2|_{input}$, in the range of $2.450 \times 10^{-3}\text{eV}^2 \leq |\Delta m^2|_{input} \leq 2.580 \times 10^{-3}\text{eV}^2$ when the 3% relative energy resolution is included. We performed additional data collection ignoring the systematic uncertainties in order to provide a strong evidence for the result. How the $|\overline{\Delta\chi^2}|$ oscillations with $\Delta m^2_{input}$ reflects on the neutrino mass hierarchy determination sensitivity depends on how the significance will be calculated, for example using Equation 6 or Equation 8.

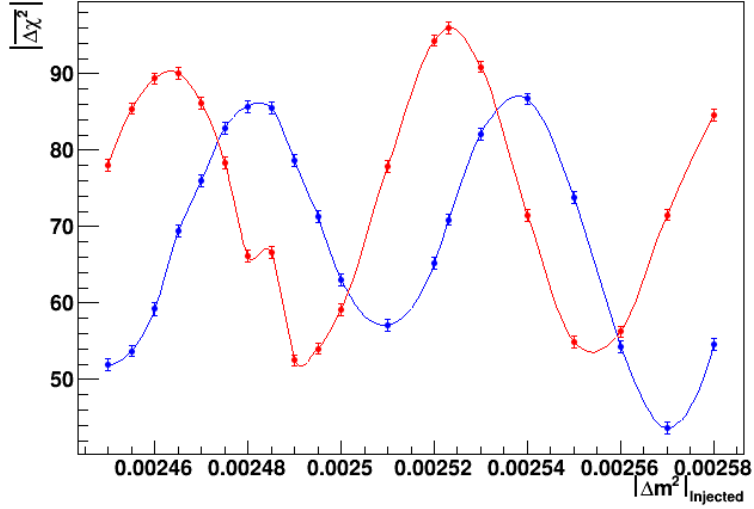

**Figure 9:** $|\overline{\Delta\chi^2}|$ variation with $|\Delta m^2|_{input}$ in the range of $2.450 \times 10^{-3}\text{eV}^2 \leq |\Delta m^2|_{input} \leq 2.580 \times 10^{-3}\text{eV}^2$ for 1000 (NH) + 1000 (IH) toy JUNO-like simulations for each point of $|\Delta m^2|_{input}$ with six years of exposure and the ten near reactor cores assuming an infinite energy resolution. The error bars correspond to the standard error of the $|\overline{\Delta\chi^2}|$ that is calculated as the standard deviation of the $\Delta\chi^2$ distribution divided by the square root of the sample size.

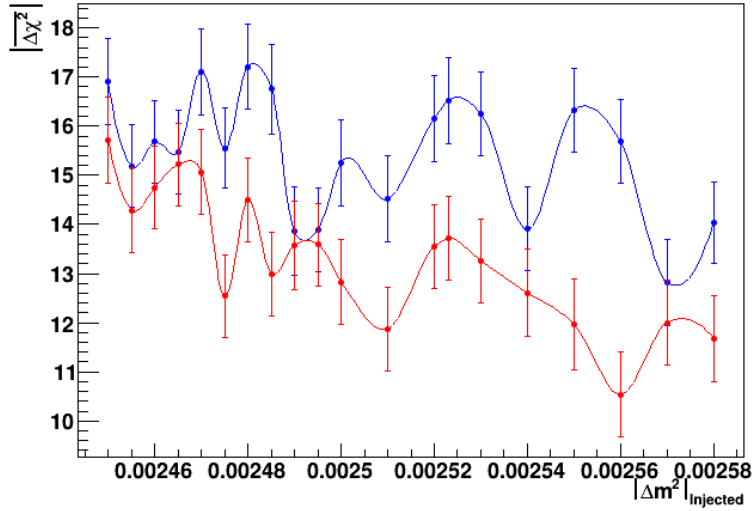

**Figure 10:** $|\overline{\Delta\chi^2}|$ varies with $|\Delta m^2|_{input}$ in the range of $2.450 \times 10^{-3}\text{eV}^2 \leq |\Delta m^2| \leq 2.580 \times 10^{-3}\text{eV}^2$ for 1000 (NH) + 1000 (IH) toy JUNO like simulations for each point of $|\Delta m^2|_{input}$ with six years of exposure and the ten near reactor cores assuming 3% relative energy resolution. The error bars correspond to the standard error of the $|\overline{\Delta\chi^2}|$ that is calculated as the standard deviation of the $\Delta\chi^2$ distribution divided by the square root of the sample size.

In case the approximation is not valid, the Z-test for 1D, Equation 6, can be used to calculate the neutrino MH sensitivity. As expected the variation of the estimator $|\overline{\Delta\chi^2}|$ will influence neutrino MH sensitivity. Figure 11 confirms the influence on neutrino MH sensitivity in case that only the statistical uncertainties are included and the sensitivity varies from about $4.5\,\sigma$ to $7.5\,\sigma$. Figure 12 confirms this influence in case that the systematic and statistical uncertainties are included and the sensitivity oscillates from about $0.9\,\sigma$ to $1.5\,\sigma$.

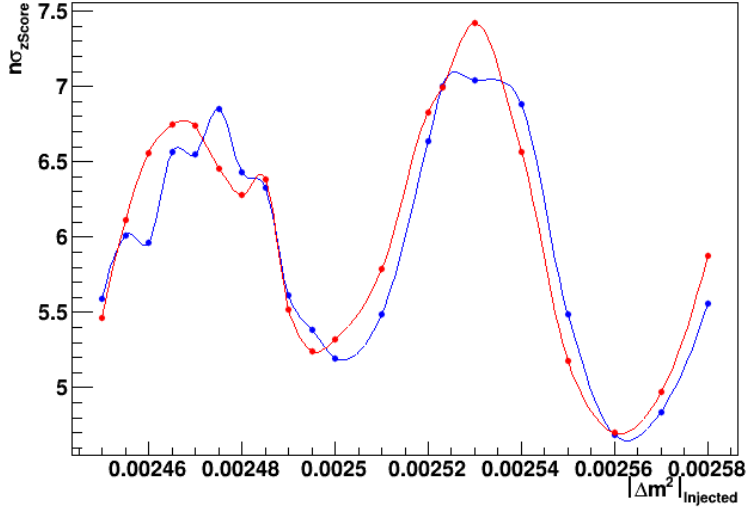

**Figure 11:** *The oscillation of significance with $|\Delta m^2|_{input}$ in the range of $2.450\times10^{-3}\mathrm{eV}^2 \leq |\Delta m^2| \leq 2.580\times10^{-3}\mathrm{eV}^2$ for 1000 (NH) + 1000 (IH) JUNO-toy like simulations for one banchmark assuming an infinite energy resolution where blue line is for NH sample and red line is for IH sample. The sensitivity using the Equation 6 varies from about $4.5\,\sigma$ to $7.5\,\sigma$.*

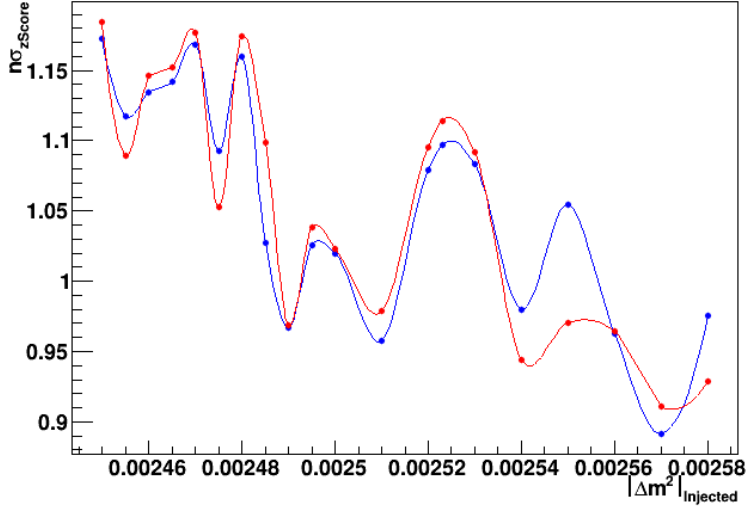

**Figure 12:** *The variation of significance with $|\Delta m^2|_{input}$ in the range of $2.450\times10^{-3}\mathrm{eV}^2 \leq |\Delta m^2| \leq 2.580\times10^{-3}\mathrm{eV}^2$ for 1000 (NH) + 1000 (IH) JUNO-toy like simulations for one banchmark assuming $3\%/\sqrt{E}$ energy resolution where blue line is for NH sample and red line is for IH sample. The sensitivity using the Equation 6 oscillates from about $0.9\,\sigma$ to $1.5\,\sigma$.*

Assuming the approximation of Equation 7 is valid at infinite energy resolution, the neutrino mass hierarchy determination sensitivity is expected to have large variation with the input parameter as confirmed in Figure 14. The sensitivity may vary from about $9.5\,\sigma$ to $7.5\,\sigma$.

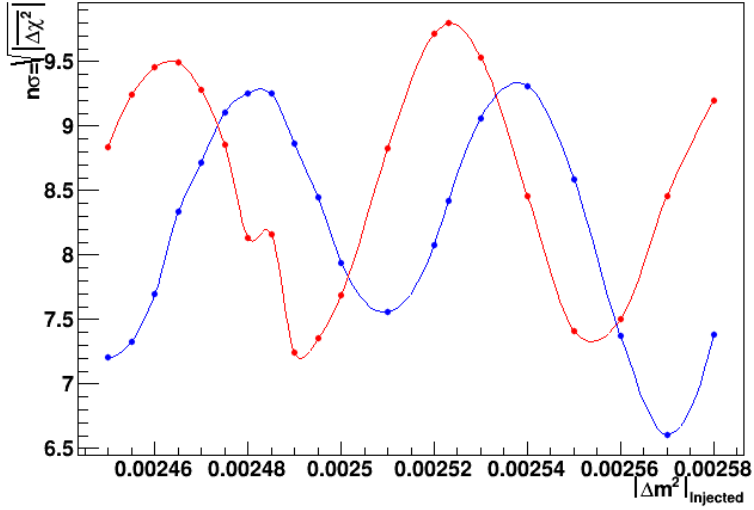

**Figure 13:** *The oscillation of significance with $|\Delta m^2|_{input}$ in the range of $2.450 \times 10^{-3}\text{eV}^2 \leq |\Delta m^2| \leq 2.580 \times 10^{-3}\text{eV}^2$ for 1000 (NH) + 1000 (IH) JUNO-toy like simulations for one banchmark assuming an infinite energy resolution where blue line is for NH sample and red line is for IH sample. The sensitivity using the Equation 8 varies from about $6.5\,\sigma$ to $9.5\,\sigma$.*

Assuming that the approximation Equation 7 is still valid at 3% relative energy resolution, the neutrino mass hierarchy determination sensitivity is not robust as confirmed in Figure 14. The sensitivity using the Equation 8 varies from a maximum of $4.1\,\sigma$ to about $3.2\,\sigma$.

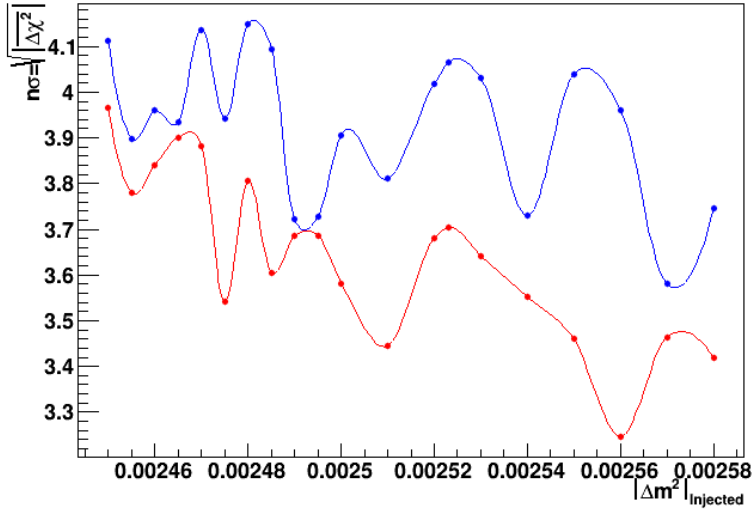

**Figure 14:** *The variation of significance with $|\Delta m^2|_{input}$ in the range of $2.450 \times 10^{-3}\text{eV}^2 \leq |\Delta m^2| \leq 2.580 \times 10^{-3}\text{eV}^2$ for 1000 (NH) + 1000 (IH) JUNO-toy like simulations for one benchmark assuming $3\%/\sqrt{E}$ energy resolution where blue line is for NH sample and red line is for IH sample. The sensitivity using the Equation 8 varies from about $3.2\,\sigma$ to $4.1\,\sigma$.*

## 5.2 The $\chi^2$ Robustness

The significance using $\chi^2$ as bi-dimensional distribution through Equation 9 varies from $1.3\,\sigma$ to $0.9\,\sigma$ assuming an infinite energy resolution as shown in Figure 15 and from $0.24\,\sigma$ to $0.18\,\sigma$ assuming 3% relative energy resolution, as shown in Figure 16.

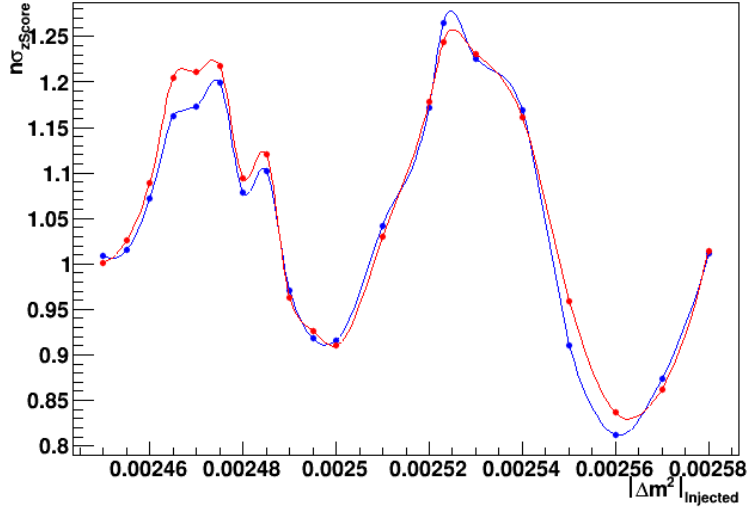

**Figure 15:** *The oscillation of significance using $\chi^2$ as bi-dimensional distribution through Equation 9 with $|\Delta m^2|_{input}$ in the range of $2.450 \times 10^{-3} \text{eV}^2 \leq |\Delta m^2|_{input} \leq 2.580 \times 10^{-3} \text{eV}^2$ for 1000 (NH) + 1000 (IH) JUNO-toy like simulations for one banchmark assuming an infinite energy resolution where blue line is for NH sample and red line is for IH sample. The significance varies from about $0.8\,\sigma$ to $1.3\,\sigma$.*

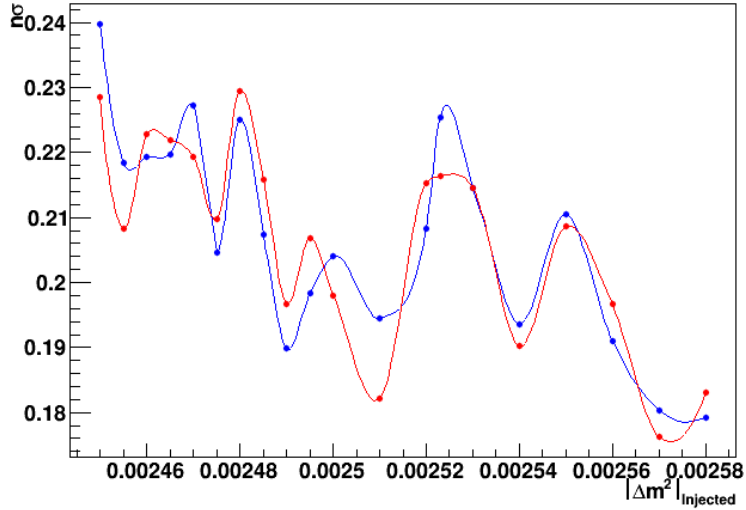

**Figure 16:** *The oscillation of the experimental significance using $\chi^2$ as bi-dimensional distribution with $|\Delta m^2|_{input}$ in the range of $2.450 \times 10^{-3} \text{eV}^2 \leq |\Delta m^2|_{input} \leq 2.580 \times 10^{-3} \text{eV}^2$ for 1000 (NH) + 1000 (IH) JUNO-toy like simulations for one banchmark assuming an 3% relative energy resolution where blue line is for NH sample and red line is for IH sample. The significance using Equation 9 varies from about $0.175\,\sigma$ to $0.24\,\sigma$.*

The oscillation of the experimental sensitivity with the value of the input parameter, the neutrino atmospheric mass difference ($|\Delta m^2|_{input}$), implies that the standard method results have strong dependency on the input parameter value. Whatever the approximation is not valid or not, systematic uncertainties included or not, this dependence still hold.

# 6 Results

In order to present the findings as clear as possible, it is imperative to study the three reported issues of the standard algorithm in the range of the atmospheric mass between $2.450 \times 10^{-3} \text{eV}^2$ and $2.580 \times 10^{-3} \text{eV}^2$. Theses issues are categorized into two types depending on which estimator being used. The first sensitivity category using $\Delta \chi^2$ estimator is reported in subsection 6.1. The second sensitivity category using $\chi^2$ is reported in subsection 6.2. For each category, a detailed study is provided for 20 different values of the atmospheric mass in the range of $2.450 \times 10^{-3} \text{eV}^2 \leq |\Delta m^2|_{input} \leq 2.580 \times 10^{-3} \text{eV}^2$, with and without systematic errors. The final results now provide solid evidences about the problematic use of the standard algorithm.

## 6.1 The Issues of $\Delta \chi^2$

Here we report two results. First, our result on the limited power of $\Delta \chi^2$ (issue one) confirming that, when systematic uncertainties are included, the approximated Equation 7 is not acceptable in the range of neutrino atmospheric mass, $2.450 \times 10^{-3} \text{eV}^2 \leq |\Delta m^2|_{input} \leq 2.580 \times 10^{-3} \text{eV}^2$. We provide the results of 20 different values of the $|\Delta m^2|_{input}$ in that range showing the limit of the approximation when including the systematic uncertainties (as confirmed in Figure 18). Although Equation 7 is widely accepted, it suffers from some limitations due to its limitation when systematic uncertainties are included (Figure 18). The limitation manifests itself decreasing the power of the $\Delta \chi^2$ estimator to determine the correct neutrino MH. The reasons behind this limitation are explained in details in section 3. As a result the power of this estimator for the MH discrimination is not promising as reported in Table 12. On the contrary, without including the systematic uncertainties Equation 7 is valid and the $\Delta \chi^2$ results are very good as reported in Figure 17 and Table 11. Second, the studies about the $\Delta \chi^2$ robustness in the range $2.450 \times 10^{-3} \text{eV}^2 \leq |\Delta m^2|_{input} \leq 2.580 \times 10^{-3} \text{eV}^2$ shows its dependence. This result is directly in line with previous result in section 5. From these sensitivity tables (Table 11 and Table 12), it is clear that the experimental sensitivity using $\Delta \chi^2$ has strongly dependence on the value of the input atmospheric mass. If the value of the input parameter, input atmospheric mass, is modified, the experimental sensitivity will change according to it. This change is not affected by the systematic uncertainties. It is an intrinsic property of the $\Delta \chi^2$ itself. Table 11 shows the sensitivities using $\Delta \chi^2$ with infinite energy resolution. As can be seen in the table, the experimental sensitivities vary a lot with different values of the neutrino atmospheric mass proving that the robustness of $\Delta \chi^2$ is not well established even at infinite energy resolution. Table 12 provides the sensitivities including the systematic uncertainties: the neutrino mass ordering discrimination varies a lot. The implications of this issue is fully discussed in section 5.

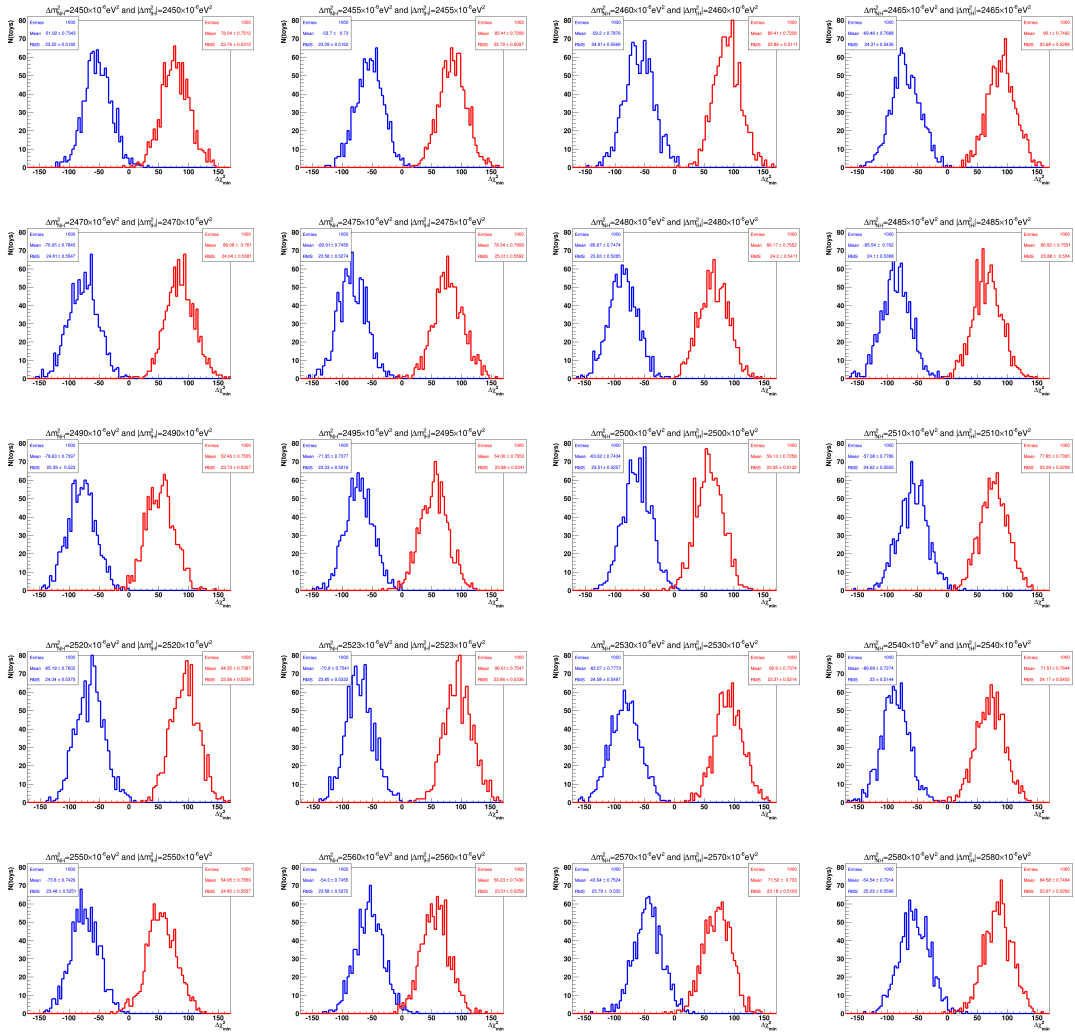

**Figure 17:** $\Delta\chi^2$ *estimator for 1000 (NH) + 1000 (IH) toy JUNO-like simulations generated at 20 different values of the atmospheric mass in the range of* $2.450 \times 10^{-3} \mathrm{eV}^2 \leq |\Delta m^2| \leq 2.580 \times 10^{-3} \mathrm{eV}^2$ *for NH hypothesis (blue distribution in each plot) and IH hypothesis (red distribution in each plot) with six years of exposure and the ten near reactor cores. An infinite energy resolution is assumed. The sensitivities due to these conditions are reported in Table 11.*

**Table 11:** *The comparison of the MH sensitivity using $\Delta\chi^2$ assuming infinite energy resolution for NH sample and IH sample, for 20 different values of the atmospheric mass in the range of $2.450\times10^{-3}\text{eV}^2 \leq |\Delta m^2| \leq 2.580\times10^{-3}\text{eV}^2$. The table indicates the sensitivity calculations using the Z-test for 1D test in two cases. The first case is without the approximation of Equation 7 and the second one is obtained using the approximation of Equation 7.*

| | Infinite Energy Resolution | | | | | | | |
|---|---|---|---|---|---|---|---|---|
| $|\Delta m^2|_{NH/IH} \times 10^{-3}$ | 2.450 | | 2.455 | | 2.460 | | 2.465 | |
| $\mu_{NH}$ | $-51.90 \pm 0.735$ | | $-53.72 \pm 0.732$ | | $-59.20 \pm 0.788$ | | $-69.43 \pm 0.7681$ | |
| $\sigma_{NH}$ | $23.24 \pm 0.520$ | | $23.14 \pm 0.518$ | | $24.91 \pm 0.557$ | | $24.29 \pm 0.5431$ | |
| $\mu_{IH}$ | $78.03 \pm 0.752$ | | $85.41 \pm 0.720$ | | $89.41 \pm 0.723$ | | $90.09 \pm 0.7482$ | |
| $\sigma_{IH}$ | $23.77 \pm 0.532$ | | $22.76 \pm 0.520$ | | $22.86 \pm 0.511$ | | $23.65 \pm 0.5291$ | |
| $z_{score}^{(NH)}$ | 5.590 | 7.204(app.) | 6.013 | 7.329(app.) | 5.966 | 7.694(app.) | 6.567 | 8.332(app.) |
| $z_{score}^{(IH)}$ | 5.466 | 8.833(app.) | 6.113 | 9.242(app.) | 6.501 | 9.456(app.) | 6.745 | 9.456(app.) |
| $|\Delta m^2|_{NH/IH} \times 10^{-3}$ | 2.470 | | 2.475 | | 2.480 | | 2.485 | |
| $\mu_{NH}$ | $-76.04 \pm 0.7834$ | | $-82.90 \pm 0.7452$ | | $-55.70 \pm 0.7471$ | | $-85.54 \pm 0.7595$ | |
| $\sigma_{NH}$ | $24.77 \pm 0.554$ | | $23.55 \pm 0.5269$ | | $23.62 \pm 0.5283$ | | $24.29 \pm 0.5431$ | |
| $\mu_{IH}$ | $86.13 \pm 0.762$ | | $78.36 \pm 0.7904$ | | $66.17 \pm 0.7649$ | | $90.09 \pm 0.7482$ | |
| $\sigma_{IH}$ | $24.07 \pm 0.5388$ | | $24.99 \pm 0.5589$ | | $24.19 \pm 0.5409$ | | $23.65 \pm 0.5291$ | |
| $z_{score}^{(NH)}$ | 6.547 | 8.720(app.) | 6.848 | 9.105(app.) | 5.160 | 7.463(app.) | 7.231 | 9.249(app.) |
| $z_{score}^{(IH)}$ | 6.737 | 9.281(app.) | 6.453 | 8.852(app.) | 5.038 | 8.134(app.) | 7.426 | 9.492(app.) |
| $|\Delta m^2|_{NH/IH} \times 10^{-3}$ | 2.490 | | 2.495 | | 2.500 | | 2.510 | |
| $\mu_{NH}$ | $-76.63 \pm 0.7387$ | | $-71.32 \pm 0.7365$ | | $-63.02 \pm 0.743$ | | $57.12 \pm 0.778$ | |
| $\sigma_{NH}$ | $23.36 \pm 0.5223$ | | $23.29 \pm 0.5208$ | | $23.51 \pm 0.526$ | | $24.60 \pm 0.550$ | |
| $\mu_{IH}$ | $52.48 \pm 0.7507$ | | $54.03 \pm 0.7557$ | | $59.13 \pm 0.726$ | | $77.89 \pm 0.738$ | |
| $\sigma_{IH}$ | $23.74 \pm 0.5308$ | | $23.90 \pm 0.5344$ | | $22.95 \pm 0.513$ | | $23.33 \pm 0.522$ | |
| $z_{score}^{(NH)}$ | 5.527 | 8.445(app.) | 5.382 | 8.445(app.) | 5.196 | 7.939(app.) | 5.488 | 7.556(app.) |
| $z_{score}^{(IH)}$ | 5.439 | 7.244(app.) | 5.280 | 7.351(app.) | 5.322 | 7.690(app.) | 5.787 | 8.826(app.) |
| $|\Delta m^2|_{NH/IH} \times 10^{-3}$ | 2.520 | | 2.523 | | 2.530 | | 2.540 | |
| $\mu_{NH}$ | $-65.19 \pm 0.760$ | | $-70.90 \pm 0.754$ | | $-82.07 \pm 0.777$ | | $-86.72 \pm 0.727$ | |
| $\sigma_{NH}$ | $24.04 \pm 0.538$ | | $23.85 \pm 0.533$ | | $24.58 \pm 0.550$ | | $23.00 \pm 0.514$ | |
| $\mu_{IH}$ | $94.35 \pm 0.739$ | | $96.01 \pm 0.755$ | | $90.90 \pm 0.737$ | | $71.51 \pm 0.762$ | |
| $\sigma_{IH}$ | $23.36 \pm 0.523$ | | $23.89 \pm 0.534$ | | $23.31 \pm 0.521$ | | $24.10 \pm 0.539$ | |
| $z_{score}^{(NH)}$ | 6.636 | 8.074(app.) | 6.998 | 8.420(app.) | 7.037 | 9.059(app.) | 6.880 | 9.312(app.) |
| $z_{score}^{(IH)}$ | 6.830 | 9.713(app.) | 6.987 | 9.798(app.) | 7.420 | 9.534(app.) | 6.566 | 8.456(app.) |
| $|\Delta m^2|_{NH/IH} \times 10^{-3}$ | 2.550 | | 2.560 | | 2.570 | | 2.580 | |
| $\mu_{NH}$ | $-73.80 \pm 0.743$ | | $-54.30 \pm 0.746$ | | $-43.64 \pm 0.752$ | | $-54.54 \pm 0.791$ | |
| $\sigma_{NH}$ | $23.48 \pm 0.525$ | | $23.58 \pm 0.527$ | | $23.79 \pm 0.532$ | | $25.03 \pm 0.560$ | |
| $\mu_{IH}$ | $54.95 \pm 0.786$ | | $56.23 \pm 0.744$ | | $71.52 \pm 0.733$ | | $84.58 \pm 0.748$ | |
| $\sigma_{IH}$ | $24.85 \pm 0.556$ | | $23.51 \pm 0.526$ | | $23.18 \pm 0.518$ | | $23.67 \pm 0.529$ | |
| $z_{score}^{(NH)}$ | 5.483 | 8.591(app.) | 4.687 | 7.369(app.) | 4.841 | 6.606(app.) | 5.848 | 7.385(app.) |
| $z_{score}^{(IH)}$ | 5.181 | 7.413(app.) | 4.701 | 7.50(app.) | 5.0 | 8.457(app.) | 5.877 | 9.197(app.) |

As mentioned in section 3, the MH sensitivities using Z-test, $z_{score}^{(NH)}$ and $z_{score}^{(IH)}$, do not exact equal to the MH sensitivities obtained in the approximated Equation 7, $z_{score}^{(NH)}(app.)$ and $z_{score}^{(IH)}(app.)$. Table 12 reports this observation for 20 different values for the atmospheric mass at infinite energy resolution providing a solid experimental evidence for over-estimation behavior for this approximation.

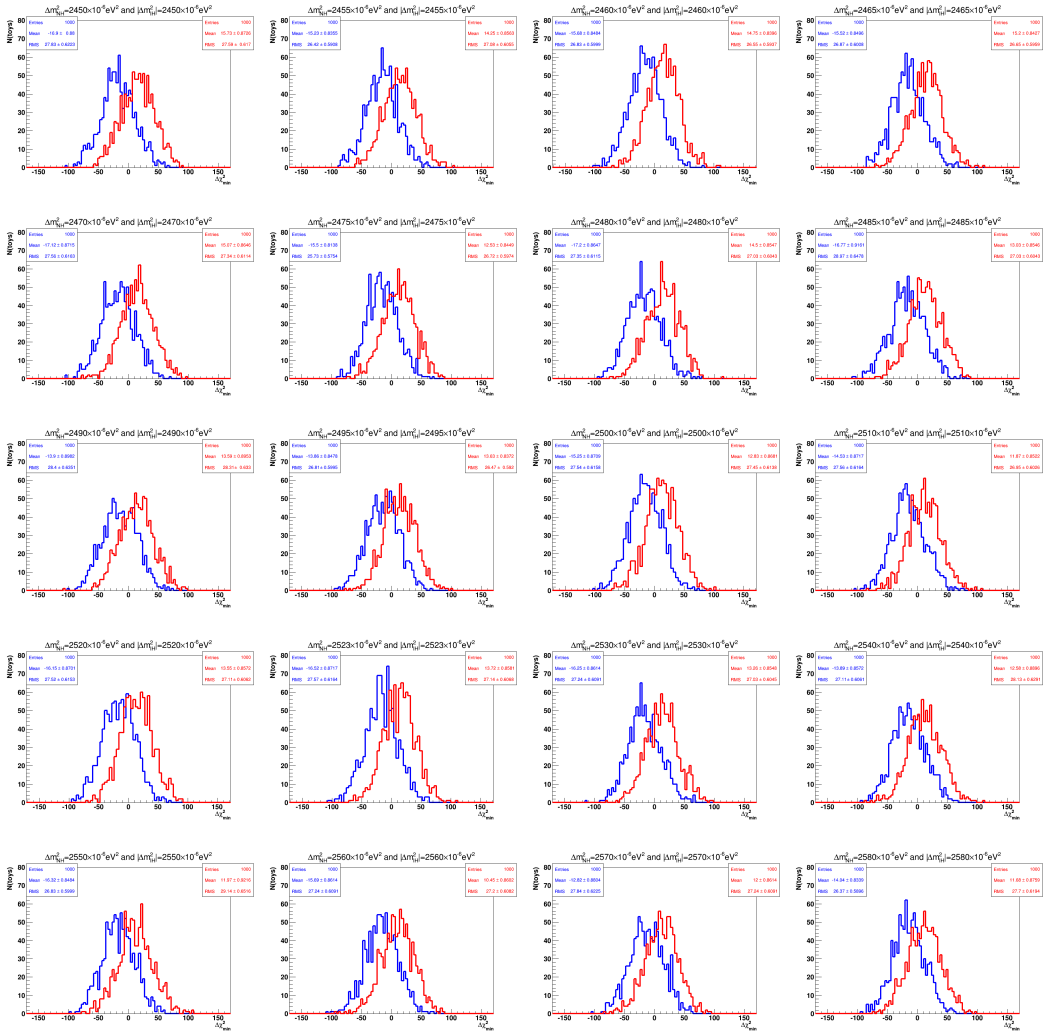

**Figure 18:** $\Delta\chi^2$ estimator for 1000 (NH) + 1000 (IH) toy JUNO-like simulations generated at 20 different values of the atmospheric mass in the range of $2.450 \times 10^{-3}\text{eV}^2 \leq |\Delta m^2| \leq 2.580 \times 10^{-3}\text{eV}^2$ for NH hypothesis (blue distribution in each plot) and IH hypothesis (red distribution in each plot) with six years of exposure and the ten near reactor cores, with energy resolution $3\%/\sqrt{E}$. The sensitivities due to these conditions are reported in Table 12.

**Table 12:** *The comparison of the MH sensitivity using $\Delta\chi^2$ for actual distributions for NH sample and IH sample, for for 20 different values of the atmospheric mass in the range of $2.450 \times 10^{-3}\mathrm{eV}^2 \leq |\Delta m^2| \leq 2.580 \times 10^{-3}\mathrm{eV}^2$. The table indicates the sensitivity calculations using the Z-test for 1D test in two cases. The first case is without the approximation of Equation 7 and the second one is using the approximation of Equation 7.*

| | Relative Energy Resolution $3\%/\sqrt{E}$ | | | | | | | |
|---|---|---|---|---|---|---|---|---|
| $|\Delta m^2|_{NH/IH} \times 10^{-3}$ | 2.450 | | 2.455 | | 2.460 | | 2.465 | |
| $\mu_{NH}$ | $-16.91 \pm 0.880$ | | $-15.19 \pm 0.834$ | | $-15.68 \pm 0.8484$ | | $-15.48 \pm 0.85$ | |
| $\sigma_{NH}$ | $27.82 \pm 0.622$ | | $26.38 \pm 0.590$ | | $26.83 \pm 0.5999$ | | $26.88 \pm 0.601$ | |
| $\mu_{IH}$ | $15.72 \pm 0.871$ | | $14.29 \pm 0.856$ | | $14.75 \pm 0.8396$ | | $15.22 \pm 0.8427$ | |
| $\sigma_{IH}$ | $27.55 \pm 0.616$ | | $27.06 \pm 0.605$ | | $26.55 \pm 0.5937$ | | $26.65 \pm 0.5959$ | |
| $z_{score}^{(NH)}$ | 1.173 | 4.112(app.) | 1.118 | 3.897(app.) | 1.134 | 3.960(app.) | 1.142 | 3.934(app.) |
| $z_{score}^{(IH)}$ | 1.184 | 3.965(app.) | 1.089 | 3.780(app.) | 1.146 | 3.841(app.) | 1.152 | 3.901(app.) |
| $|\Delta m^2|_{NH/IH} \times 10^{-3}$ | 2.470 | | 2.475 | | 2.480 | | 2.485 | |
| $\mu_{NH}$ | $-17.10 \pm 0.8709$ | | $-15.55 \pm 0.8126$ | | $-17.21 \pm 0.8646$ | | $-16.76 \pm 0.9159$ | |
| $\sigma_{NH}$ | $27.54 \pm 0.6158$ | | $25.70 \pm 0.5746$ | | $27.34 \pm 0.6114$ | | $28.96 \pm 0.6477$ | |
| $\mu_{IH}$ | $15.07 \pm 0.8645$ | | $12.54 \pm 0.8437$ | | $14.49 \pm 0.8539$ | | $12.99 \pm 0.856$ | |
| $\sigma_{IH}$ | $27.34 \pm 0.6113$ | | $26.68 \pm 0.5966$ | | $27.00 \pm 0.6038$ | | $27.07 \pm 0.6053$ | |
| $z_{score}^{(NH)}$ | 1.168 | 4.135(app.) | 1.093 | 3.943(app.) | 1.159 | 4.148(app.) | 1.027 | 4.094(app.) |
| $z_{score}^{(IH)}$ | 1.177 | 3.882(app.) | 1.053 | 3.541(app.) | 1.174 | 3.807(app.) | 1.099 | 3.604(app.) |
| $|\Delta m^2|_{NH/IH} \times 10^{-3}$ | 2.490 | | 2.495 | | 2.500 | | 2.510 | |
| $\mu_{NH}$ | $-13.86 \pm 0.8974$ | | $-13.89 \pm 0.8476$ | | $-15.25 \pm 0.8709$ | | $14.52 \pm 0.871$ | |
| $\sigma_{NH}$ | $28.38 \pm 0.6345$ | | $26.80 \pm 0.5994$ | | $27.54 \pm 0.6158$ | | $27.55 \pm 0.616$ | |
| $\mu_{IH}$ | $13.58 \pm 0.8955$ | | $13.59 \pm 0.8372$ | | $12.83 \pm 0.8681$ | | $11.87 \pm 0.853$ | |
| $\sigma_{IH}$ | $28.32 \pm 0.6332$ | | $26.47 \pm 0.5920$ | | $27.45 \pm 0.6138$ | | $26.97 \pm 0.603$ | |
| $z_{score}^{(NH)}$ | 0.967 | 3.723(app.) | 1.025 | 3.727(app.) | 1.020 | 3.905(app.) | 0.958 | 3.811(app.) |
| $z_{score}^{(IH)}$ | 0.969 | 3.685(app.) | 1.038 | 3.686(app.) | 1.023 | 3.582(app.) | 0.978 | 3.445(app.) |
| $|\Delta m^2|_{NH/IH} \times 10^{-3}$ | 2.520 | | 2.523 | | 2.530 | | 2.540 | |
| $\mu_{NH}$ | $-16.15 \pm 0.870$ | | $-16.52 \pm 0.872$ | | $-16.25 \pm 0.861$ | | $-13.91 \pm 0.856$ | |
| $\sigma_{NH}$ | $27.52 \pm 0.615$ | | $27.57 \pm 0.616$ | | $27.24 \pm 0.609$ | | $27.07 \pm 0.605$ | |
| $\mu_{IH}$ | $13.55 \pm 0.857$ | | $13.72 \pm 0.858$ | | $13.26 \pm 0.855$ | | $12.61 \pm 0.888$ | |
| $\sigma_{IH}$ | $27.11 \pm 0.606$ | | $27.14 \pm 0.607$ | | $27.03 \pm 0.605$ | | $28.08 \pm 0.628$ | |
| $z_{score}^{(NH)}$ | 1.079 | 4.019(app.) | 1.097 | 4.064(app.) | 1.083 | 4.031(app.) | 0.9797 | 3.30(app.) |
| $z_{score}^{(IH)}$ | 1.096 | 3.681(app.) | 1.114 | 3.704(app.) | 1.092 | 3.641(app.) | 0.944 | 3.551(app.) |
| $|\Delta m^2|_{NH/IH} \times 10^{-3}$ | 2.550 | | 2.560 | | 2.570 | | 2.580 | |
| $\mu_{NH}$ | $-16.32 \pm 0.848$ | | $-15.69 \pm 0.861$ | | $-12.82 \pm 0.880$ | | $-14.04 \pm 0.834$ | |
| $\sigma_{NH}$ | $26.83 \pm 0.600$ | | $27.24 \pm 0.609$ | | $27.84 \pm 0.623$ | | $26.37 \pm 0.590$ | |
| $\mu_{IH}$ | $11.97 \pm 0.922$ | | $10.54 \pm 0.860$ | | $12.00 \pm 0.861$ | | $11.68 \pm 0.876$ | |
| $\sigma_{IH}$ | $29.14 \pm 0.652$ | | $27.20 \pm 0.608$ | | $27.24 \pm 0.609$ | | $27.70 \pm 0.619$ | |
| $z_{score}^{(NH)}$ | 1.054 | 4.040(app.) | 0.963 | 3.961(app.) | 0.892 | 3.581(app.) | 0.975 | 3.747(app.) |
| $z_{score}^{(IH)}$ | 0.971 | 3.460(app.) | 0.964 | 3.247(app.) | 0.911 | 3.464(app.) | 0.944 | 3.418(app.) |

## 6.2 The Issues of $\chi^2$

Each plot of Figure 19 and Figure 20 proves that $\chi^2$ has not enough ability to produce high sensitivity to distinguish between the right and wrong ordering of the neutrino using the medium baseline reactor spectrum. From the sensitivity tables (Table 13 and Table 14), it is clear that the experimental sensitivity using the $\chi^2$ estimator has strongly dependence on the value of the neutrino atmospheric mass. If the neutrino atmospheric mass value is modified, the experimental sensitivity will change according to it, even when the systematic uncertainties are not included.

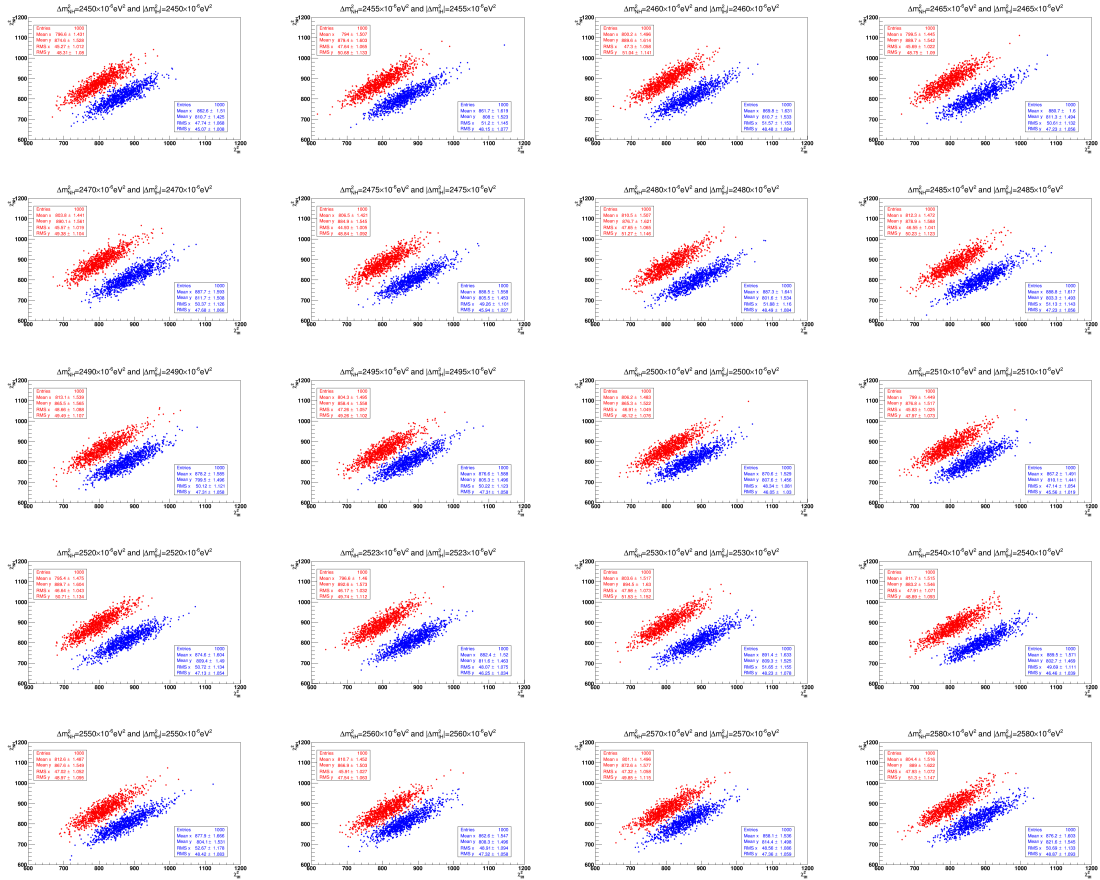

**Figure 19:** *Two $\chi^2$ distributions for 1000 (NH) + 1000 (IH) toy JUNO-like simulations that generated at 20 different values of the atmospheric mass in the range of $2.450 \times 10^{-3}\text{eV}^2 \leq |\Delta m^2| \leq 2.580 \times 10^{-3}\text{eV}^2$ for NH hypothesis (blue distribution in each plot) and IH hypothesis (blue distribution in each plot) with six years of exposure and the ten near reactor cores with infinite energy resolution. The sensitivities due to these conditions are reported in Table 13*

**Table 13:** The comparison of the MH sensitivity using $\chi^2$ as a bi-dimensional estimator assuming infinite energy resolution for, NH sample and IH sample, for 20 different values of the atmospheric mass in the range of $2.450 \times 10^{-3} \text{eV}^2 \leq |\Delta m^2| \leq 2.580 \times 10^{-3} \text{eV}^2$. The table indicates the sensitivity calculations using the Z-test for 2D.

| | Infinite Energy Resolution | | | | | | | | | |
|---|---|---|---|---|---|---|---|---|---|---|
| $|\Delta m^2|_{NH/IH} \times 10^{-3}$ | 2.450 | | 2.455 | | 2.460 | | 2.465 | | 2.470 | |
| | NH | IH | NH | IH | NH | IH | NH | IH | NH | IH |
| $\mu_{NH}$ | 810.7 ± 1.425 | 874.6 ± 1.528 | 808 ± 1.523 | 879.4 ± 1.602 | 810.7 ± 1.533 | 889.6 ± 1.614 | 811.3 ± 1.494 | 889.7 ± 1.542 | 811.7 ± 1.593 | 890.1 ± 1.441 |
| $\sigma_{NH}$ | 45.07 ± 1.008 | 48.31 ± 1.080 | 48.17 ± 1.077 | 50.67 ± 1.133 | 48.48 ± 1.084 | 51.04 ± 1.141 | 47.23 ± 1.056 | 48.76 ± 1.090 | 47.7 ± 1.067 | 49.39 ± 1.104 |
| $\mu_{IH}$ | 862.6 ± 1.510 | 796.6 ± 1.432 | 861.7 ± 1.619 | 794 ± 1.507 | 869.8 ± 1.631 | 800.2 ± 1.496 | 880.7 ± 1.601 | 799.5 ± 1.445 | 887.7 ± 1.593 | 803.8 ± 1.441 |
| $\sigma_{IH}$ | 47.75 ± 1.068 | 45.27 ± 1.012 | 51.2 ± 1.145 | 47.67 ± 1.066 | 51.57 ± 1.153 | 47.3 ± 1.058 | 50.62 ± 1.132 | 45.69 ± 1.022 | 50.39 ± 1.127 | 45.57 ± 1.019 |
| $z_{score}^{(NH)}$ | 1.0095 | | 1.0154 | | 1.0717 | | 1.163 | | 1.1728 | |
| $z_{score}^{(IH)}$ | 1.0013 | | 1.0261 | | 1.0891 | | 1.2050 | | 1.2113 | |
| $|\Delta m^2|_{NH/IH} \times 10^{-3}$ | 2.475 | | 2.480 | | 2.485 | | 2.490 | | 2.495 | |
| $\mu_{NH}$ | 811.7 ± 1.508 | 890.1 ± 1.562 | 805.5 ± 1.453 | 884.9 ± 1.545 | 801.3 ± 1.641 | 876.7 ± 1.622 | 799.6 ± 1.496 | 865.5 ± 1.565 | 805.3 ± 1.495 | 858.4 ± 1.558 |
| $\sigma_{NH}$ | 47.7 ± 1.067 | 49.39 ± 1.104 | 45.94 ± 1.027 | 48.85 ± 1.092 | 48.48 ± 1.084 | 51.28 ± 1.147 | 47.32 ± 1.058 | 49.50 ± 1.107 | 47.28 ± 1.057 | 49.28 ± 1.102 |
| $\mu_{IH}$ | 887.7 ± 1.593 | 803.8 ± 1.441 | 888.5 ± 1.558 | 806.5 ± 1.420 | 887.3 ± 1.641 | 810.5 ± 1.507 | 878.2 ± 1.585 | 813.10 ± 1.539 | 876.6 ± 1.588 | 804.3 ± 1.495 |
| $\sigma_{IH}$ | 50.39 ± 1.127 | 45.57 ± 1.019 | 49.26 ± 1.102 | 44.92 | 51.9 ± 1.084 | 47.64 ± 1.065 | 50.11 ± 1.121 | 48.66 ± 1.088 | 47.28 ± 1.057 | 47.27 ± 1.057 |
| $z_{score}^{(NH)}$ | 1.1990 | | 1.0786 | | 1.102 | | 0.9703 | | 0.9177 | |
| $z_{score}^{(IH)}$ | 1.2172 | | 1.0944 | | 1.1201 | | 0.9632 | | 0.9268 | |
| $|\Delta m^2|_{NH/IH} \times 10^{-3}$ | 2.500 | | 2.510 | | 2.520 | | 2.523 | | 2.530 | |
| $\mu_{NH}$ | 807.6 ± 1.456 | 865.3 ± 1.522 | 810.10 ± 1.491 | 876.90 ± 1.516 | 809.4 ± 1.49 | 889.7 ± 1.604 | 811.6 ± 1.463 | 892.6 ± 1.573 | 809.3 ± 1.525 | 894.5 ± 1.630 |
| $\sigma_{NH}$ | 46.05 ± 1.03 | 48.12 ± 1.076 | 45.57 ± 1.019 | 47.95 ± 1.072 | 47.13 ± 1.054 | 50.71 ± 1.134 | 46.25 ± 1.034 | 49.74 ± 1.112 | 48.23 ± 1.078 | 51.53 ± 1.152 |
| $\mu_{IH}$ | 870.6 ± 1.529 | 806.2 ± 1.483 | 867.2 ± 1.491 | 799 ± 1.449 | 874.6 ± 1.604 | 795.4 ± 1.475 | 882.4 ± 1.52 | 796.6 ± 1.460 | 891.4 ± 1.633 | 803.6 ± 1.517 |
| $\sigma_{IH}$ | 48.34 ± 1.081 | 46.91 ± 1.049 | 47.14 ± 1.054 | 45.83 ± 1.516 | 50.72 ± 1.134 | 46.64 ± 1.043 | 48.07 ± 1.075 | 46.17 ± 1.032 | 51.65 ± 1.155 | 51.53 ± 1.152 |
| $z_{score}^{(NH)}$ | 0.9159 | | 1.0418 | | 1.1720 | | 1.2647 | | 1.226 | |
| $z_{score}^{(IH)}$ | 0.9099 | | 1.0200 | | 1.1781 | | 1.2442 | | 1.231 | |
| $|\Delta m^2|_{NH/IH} \times 10^{-3}$ | 2.540 | | 2.550 | | 2.560 | | 2.570 | | 2.580 | |
| $\mu_{NH}$ | 802.7 ± 1.469 | 883.2 ± 1.546 | 804.1 ± 1.531 | 867.6 ± 1.549 | 862.6 ± 1.547 | 866.9 ± 1.503 | 814.4 ± 1.498 | 872.6 ± 1.577 | 821.6 ± 1.545 | 889 ± 1.622 |
| $\sigma_{NH}$ | 46.47 ± 1.039 | 48.89 ± 1.093 | 48.42 ± 1.083 | 48.97 ± 1.095 | 47.32 ± 1.058 | 47.54 ± 1.063 | 47.36 ± 1.059 | 49.85 ± 1.115 | 48.87 ± 1.093 | 51.3 ± 1.147 |
| $\mu_{IH}$ | 889.5 ± 1.571 | 811.7 ± 1.515 | 877.90 ± 1.666 | 812.6 ± 1.487 | 808.3 ± 1.496 | 810.7 ± 1.452 | 858.1 ± 1.536 | 801.1 ± 1.496 | 876.2 ± 1.603 | 804.4 ± 1.516 |
| $\sigma_{IH}$ | 49.68 ± 1.111 | 47.91 ± 1.071 | 48.42 ± 1.083 | 47.02 ± 1.052 | 48.91 ± 1.058 | 45.91 ± 1.027 | 48.56 ± 1.086 | 47.32 ± 1.058 | 50.69 ± 1.133 | 47.93 ± 1.072 |
| $z_{score}^{(NH)}$ | 1.1690 | | 0.9103 | | 0.8122 | | 0.8734 | | 1.0111 | |
| $z_{score}^{(IH)}$ | 1.1613 | | 0.9586 | | 0.8364 | | 0.8622 | | 1.0144 | |

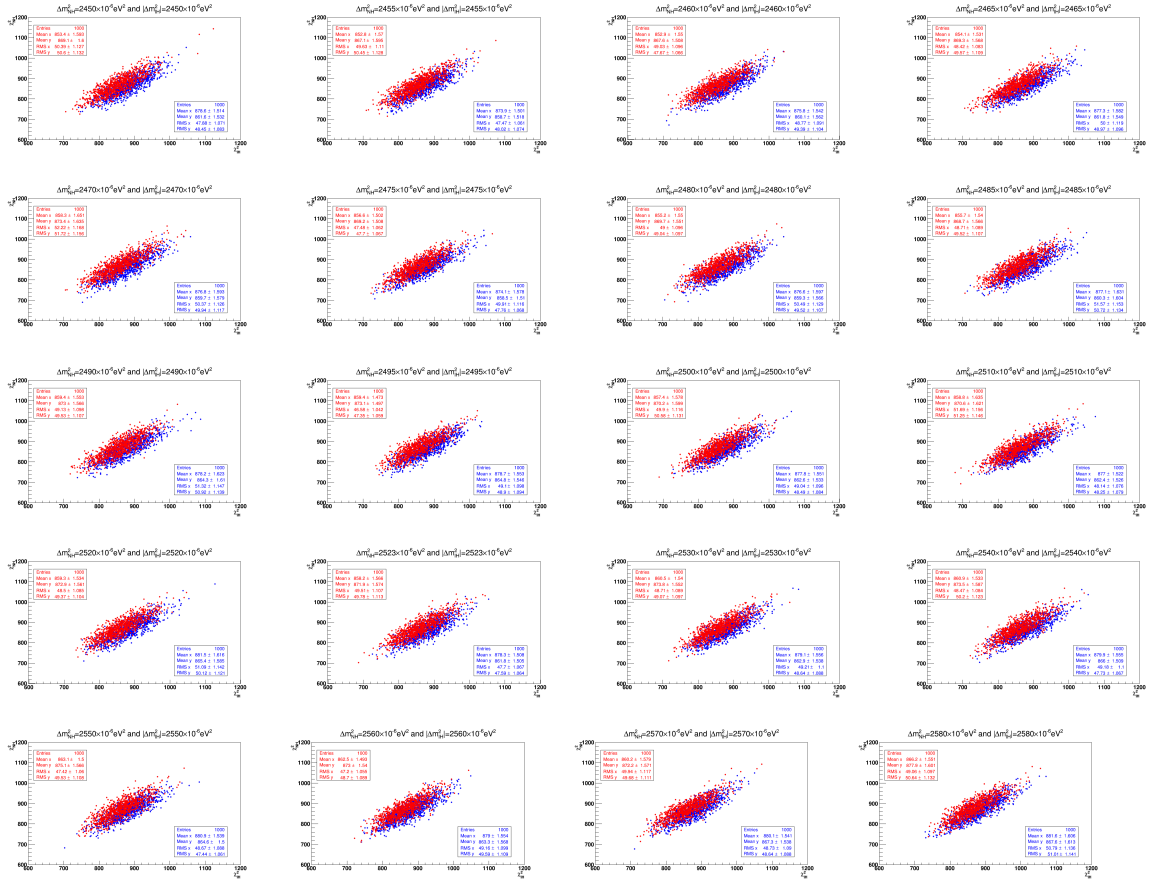

**Figure 20:** *Two $\chi^2$ distributions for 1000 (NH) + 1000 (IH) toy JUNO-like simulations generated at 20 different values of the atmospheric mass in the range of $2.450 \times 10^{-3} \text{eV}^2 \leq |\Delta m^2|_{input} \leq 2.580 \times 10^{-3} \text{eV}^2$ for NH hypothesis (blue distribution in each plot) and IH hypothesis (blue distribution in each plot) with six years of exposure and the ten near reactor cores, with energy resolution $3\%/\sqrt{E}$. The sensitivities due to these conditions are reported in Table 14.*

The results about the standard algorithm confirmed the three statistical issues in the range $2.450 \times 10^{-3} \text{eV}^2 \leq |\Delta m^2|_{input} \leq 2.580 \times 10^{-3} \text{eV}^2$.

# 7  Conclusion

Advances in statistical methods may play a decisive role in the discovery reached at neutrino physics experiments. Evaluating the used statistical methods and updating them is a necessary step in building a robust statistical analysis for answering the open questions in neutrino physics [20]. The statistical issues on the $\nu$ MHD from the reactor experiments have been illustrated, starting from the limited power of the $\Delta \chi^2$. When the simulation is performed on an event-by-event basis and not on a semi-analytical one, the significance drastically drops. In fact, the systematic uncertainties due to the 3% relatively energy resolution causes unbalanced migration effect between events that do not show up except the simulations are done on a event-by-event basis. To confirm the effect, the simulations at infinite energy resolution are done supporting the validation of the assumption of Equation 7 in case of exclusion of the systematic uncertainties. $\Delta \chi^2$ is fully controlled by the statistical assumptions as explained in section 3. That is the major limit to the approximation, reducing the experimental standard sensitivity that is officially reported. To conclude the first issue, the $\Delta \chi^2$ estimator provides us with different results due to different simulation procedures. Second, the strong positive correlations between the $\chi^2_{min(NH)}$ and $\chi^2_{min(IH)}$ when they are drawn in a 2 dimensional map confirms the $\chi^2 = (\chi^2_{min(IH)}, \chi^2_{min(NH)})$ being a bi-dimensional estimator. To conclude the second issue, JUNO sensitivity using $\chi^2$ as bi-dimensional estimator is not promising as well. Third, the $\Delta\chi^2$ is dominated by the $|\Delta m^2|_{input}$ value as described in section 5. Then, the MH significance using $|\overline{\Delta\chi^2}|$ depends on the values of the input parameter $|\Delta m^2|_{input}$. That is the reason we were interested in study the MHD problem using the standard method at 20 different values of $|\Delta m^2|_{input}$ in the range between $2.450 \times 10^{-3} \text{eV}^2$ and $2.580 \times 10^{-3} \text{eV}^2$.

**Table 14:** The comparison of the MH sensitivity using $\chi^2$ as a bi-dimensional estimator for actual distributions for NH sample and IH sample, for 20 different values of the atmospheric mass in the range of $2.450 \times 10^{-3}\text{eV}^2 \leq |\Delta m^2|_{input} \leq 2.580 \times 10^{-3}\text{eV}^2$. The table indicates the sensitivity calculations using the Z-test for 2D test.

| | | | | | relative energy resolution $3\%/\sqrt{E}$ | | | | | |
|---|---|---|---|---|---|---|---|---|---|---|
| $|\Delta m^2|_{NH/IH} \times 10^{-3}$ | 2.450 | | 2.455 | | 2.460 | | 2.465 | | 2.470 | |
| | NH | IH | NH | IH | NH | IH | NH | IH | NH | IH |
| $\mu_{NH}$ | $861.7 \pm 1.532$ | $869.1 \pm 1.601$ | $858.7 \pm 1.519$ | $867.1 \pm 1.596$ | $860.1 \pm 1.542$ | $867.6 \pm 1.508$ | $861.8 \pm 1.549$ | $869.4 \pm 1.567$ | $864.4 \pm 1.61$ | $873 \pm 1.566$ |
| $\sigma_{NH}$ | $48.45 \pm 1.083$ | $50.61 \pm 1.127$ | $48.03 \pm 1.074$ | $50.46 \pm 1.128$ | $49.39 \pm 1.104$ | $47.67 \pm 1.066$ | $48.97 \pm 1.096$ | $49.57 \pm 1.108$ | $50.90 \pm 1.138$ | $49.54 \pm 1.108$ |
| $\mu_{IH}$ | $878.6 \pm 1.514$ | $853.4 \pm 1.594$ | $873.8 \pm 1.501$ | $852.8 \pm 1.570$ | $875.8 \pm 1.542$ | $852.9 \pm 1.55$ | $877.3 \pm 1.582$ | $854.1 \pm 1.531$ | $878.2 \pm 1.623$ | $859.4 \pm 1.554$ |
| $\sigma_{IH}$ | $48.45 \pm 1.083$ | $50.41 \pm 1.132$ | $47.48 \pm 1.062$ | $49.64 \pm 1.110$ | $48.77 \pm 1.091$ | $49.03 \pm 1.096$ | $50 \pm 1.119$ | $49.57 \pm 1.108$ | $51.31 \pm 1.147$ | $49.14 \pm 1.099$ |
| $z_{score}^{(NH)}$ | 0.2397 | | 0.2184 | | 0.2194 | | 0.2198 | | 0.2273 | |
| $z_{score}^{(IH)}$ | 0.2286 | | 0.2084 | | 0.2227 | | 0.2220 | | 0.2193 | |
| $|\Delta m^2|_{NH/IH} \times 10^{-3}$ | 2.475 | | 2.480 | | 2.485 | | 2.490 | | 2.495 | |
| | NH | IH | NH | IH | NH | IH | NH | IH | NH | IH |
| $\mu_{NH}$ | $858.5 \pm 1.511$ | $869.2 \pm 1.509$ | $859.4 \pm 1.566$ | $869.7 \pm 1.551$ | $860.3 \pm 1.604$ | $868.7 \pm 1.565$ | $864.4 \pm 1.61$ | $873 \pm 1.566$ | $864.8 \pm 1.546$ | $873 \pm 1.497$ |
| $\sigma_{NH}$ | $47.77 \pm 1.068$ | $47.73 \pm 1.067$ | $49.51 \pm 1.107$ | $49.05 \pm 1.097$ | $50.71 \pm 1.134$ | $49.50 \pm 1.107$ | $50.9 \pm 1.138$ | $49.54 \pm 1.108$ | $48.89 \pm 1.093$ | $47.35 \pm 1.059$ |
| $\mu_{IH}$ | $874.1 \pm 1.578$ | $856.7 \pm 1.502$ | $876.6 \pm 1.596$ | $855.2 \pm 1.55$ | $877.1 \pm 1.631$ | $855.7 \pm 1.54$ | $878.2 \pm 1.623$ | $859.4 \pm 1.554$ | $878.7 \pm 1.553$ | $859.4 \pm 1.473$ |
| $\sigma_{IH}$ | $49.89 \pm 1.116$ | $47.49 \pm 1.062$ | $50.46 \pm 1.128$ | $49.05 \pm 1.097$ | $51.57 \pm 1.153$ | $48.71 \pm 1.089$ | $51.31 \pm 1.147$ | $49.14 \pm 1.099$ | $49.10 \pm 1.098$ | $46.6 \pm 1.059$ |
| $z_{score}^{(NH)}$ | 0.2045 | | 0.2250 | | 0.2073 | | 0.1898 | | 0.1983 | |
| $z_{score}^{(IH)}$ | 0.2098 | | 0.2294 | | 0.2159 | | 0.1967 | | 0.2068 | |
| $|\Delta m^2|_{NH/IH} \times 10^{-3}$ | 2.500 | | 2.510 | | 2.520 | | 2.523 | | 2.530 | |
| | NH | IH | NH | IH | NH | IH | NH | IH | NH | IH |
| $\mu_{NH}$ | $862.6 \pm 1.533$ | $870.2 \pm 1.599$ | $862.4 \pm 1.526$ | $870.6 \pm 1.621$ | $865.4 \pm 1.585$ | $872.9 \pm 1.561$ | $861.8 \pm 1.505$ | $871.9 \pm 1.574$ | $862.9 \pm 1.538$ | $873.8 \pm 1.552$ |
| $\sigma_{NH}$ | $48.49 \pm 1.084$ | $50.58 \pm 1.131$ | $48.24 \pm 1.079$ | $51.68 \pm 1.156$ | $50.12 \pm 1.121$ | $49.37 \pm 1.104$ | $47.59 \pm 1.064$ | $49.78 \pm 1.113$ | $48.64 \pm 1.088$ | $49.07 \pm 1.097$ |
| $\mu_{IH}$ | $877.8 \pm 1.551$ | $857.4 \pm 1.578$ | $877 \pm 152$ | $858.8 \pm 1.634$ | $881.5 \pm 1.616$ | $859.3 \pm 1.534$ | $878.3 \pm 1.508$ | $858.2 \pm 1.566$ | $879.1 \pm 1.556$ | $860.5 \pm 1.54$ |
| $\sigma_{IH}$ | $49.04 \pm 1.096$ | $49.90 \pm 1.116$ | $48.14 \pm 1.08$ | $51.25 \pm 1.146$ | $51.09 \pm 1.142$ | $48.50 \pm 1.065$ | $47.7 \pm 1.067$ | $49.51 \pm 1.107$ | $49.21 \pm 1.10$ | $860.5 \pm 1.54$ |
| $z_{score}^{(NH)}$ | 0.2040 | | 0.1945 | | 0.2083 | | 0.2253 | | 0.2145 | |
| $z_{score}^{(IH)}$ | 0.1980 | | 0.1822 | | 0.2153 | | 0.2164 | | 0.2146 | |
| $|\Delta m^2|_{NH/IH} \times 10^{-3}$ | 2.540 | | 2.550 | | 2.560 | | 2.570 | | 2.580 | |
| | NH | IH | NH | IH | NH | IH | NH | IH | NH | IH |
| $\mu_{NH}$ | $866.0 \pm 1.509$ | $873.5 \pm 1.587$ | $864.6 \pm 1.50$ | $875.1 \pm 1.566$ | $863.3 \pm 1.568$ | $873.0 \pm 1.540$ | $867.3 \pm 1.538$ | $872.2 \pm 1.571$ | $867.6 \pm 1.613$ | $877.9 \pm 1.601$ |
| $\sigma_{NH}$ | $47.73 \pm 1.067$ | $50.19 \pm 1.122$ | $47.44 \pm 1.061$ | $49.53 \pm 1.108$ | $49.59 \pm 1.109$ | $48.70 \pm 1.089$ | $48.64 \pm 1.088$ | $49.68 \pm 1.111$ | $51.01 \pm 1.141$ | $50.64 \pm 1.132$ |
| $\mu_{IH}$ | $879.9 \pm 1.555$ | $860.9 \pm 1.533$ | $880.9 \pm 1539$ | $863.1 \pm 1.5$ | $879.0 \pm 1.554$ | $862.5 \pm 1.493$ | $880.1 \pm 1.541$ | $860.2 \pm 1.579$ | $881.6 \pm 1.606$ | $866.20 \pm 1.551$ |
| $\sigma_{IH}$ | $49.17 \pm 1.067$ | $48.47 \pm 1.084$ | $48.67 \pm 1.061$ | $47.42 \pm 1.06$ | $49.16 \pm 1.099$ | $47.2 \pm 1.055$ | $48.73 \pm 1.09$ | $49.941.117$ | $50.79 \pm 1.136$ | $49.06 \pm 1.097$ |
| $z_{score}^{(NH)}$ | 0.1936 | | 0.2104 | | 0.1909 | | 0.1803 | | 0.1793 | |
| $z_{score}^{(IH)}$ | 0.1902 | | 0.2086 | | 0.1966 | | 0.1763 | | 0.1831 | |

# 8 Acknowledgements

We are particularly grateful to Xuefeng Ding for valuable suggestions on the robustness of $\Delta\chi^2$.

# 9 Appendix

## A Fitting with TMinuit class

A toy simulations were based on a single event basis and the expected systematic errors via a Gaussian distribution centered at the expected mean and with the standard deviation of the estimated uncertainty can be added. For JUNO, a global $3\%/\sqrt{E}$ (MeV) resolution on the energy reconstruction is expected. The oscillation parameters have been taken from the most recent global fits listed in Table 1. The Poisson statistical fluctuation is automatically included.

The fitting procedures and the minimization of $\chi^2$ are done via the ROOT minimization libraries (the TMinuit algorithm). In the minimization procedure all the oscillation parameters were fixed to the best-fitting values of [8]. A total of 108357 signal events are processed for each toy-simulations. The official version of JUNO Software "J17v1r1" is used. $\Delta\chi^2$ will be often scaled with the number of degrees of freedom, which is clearly equal to the number of fitted data minus the constraints: $bin - 6$. Figure 21 and Figure 22 indicate $\chi^2$ distributions for 1000 toy JUNO-like simulations generated for NH and IH samples respectively. The simulations are generated at 20 different values of the atmospheric mass in the range of $2.450 \times 10^{-3} \text{eV}^2 \leq |\Delta m^2|_{input} \leq 2.580 \times 10^{-3} \text{eV}^2$ for NH hypothesis (blue graphs) and for IH hypothesis (red graphs) with six years of exposure and the ten near reactor cores with an infinite energy resolution. Figure 23 and Figure 24 indicate the $\chi^2$ distributions for 1000 toy JUNO-like simulations generated for NH and IH samples respectively. The simulations are generated at 20 different values of the atmospheric mass in the range of $2.450 \times 10^{-3} \text{eV}^2 \leq |\Delta m^2|_{input} \leq 2.580 \times 10^{-3} \text{eV}^2$ for NH hypothesis (blue graphs) and for IH hypothesis (red graphs) with six years of exposure and the ten near reactor cores with an $3\%/\sqrt{E}$ energy resolution.

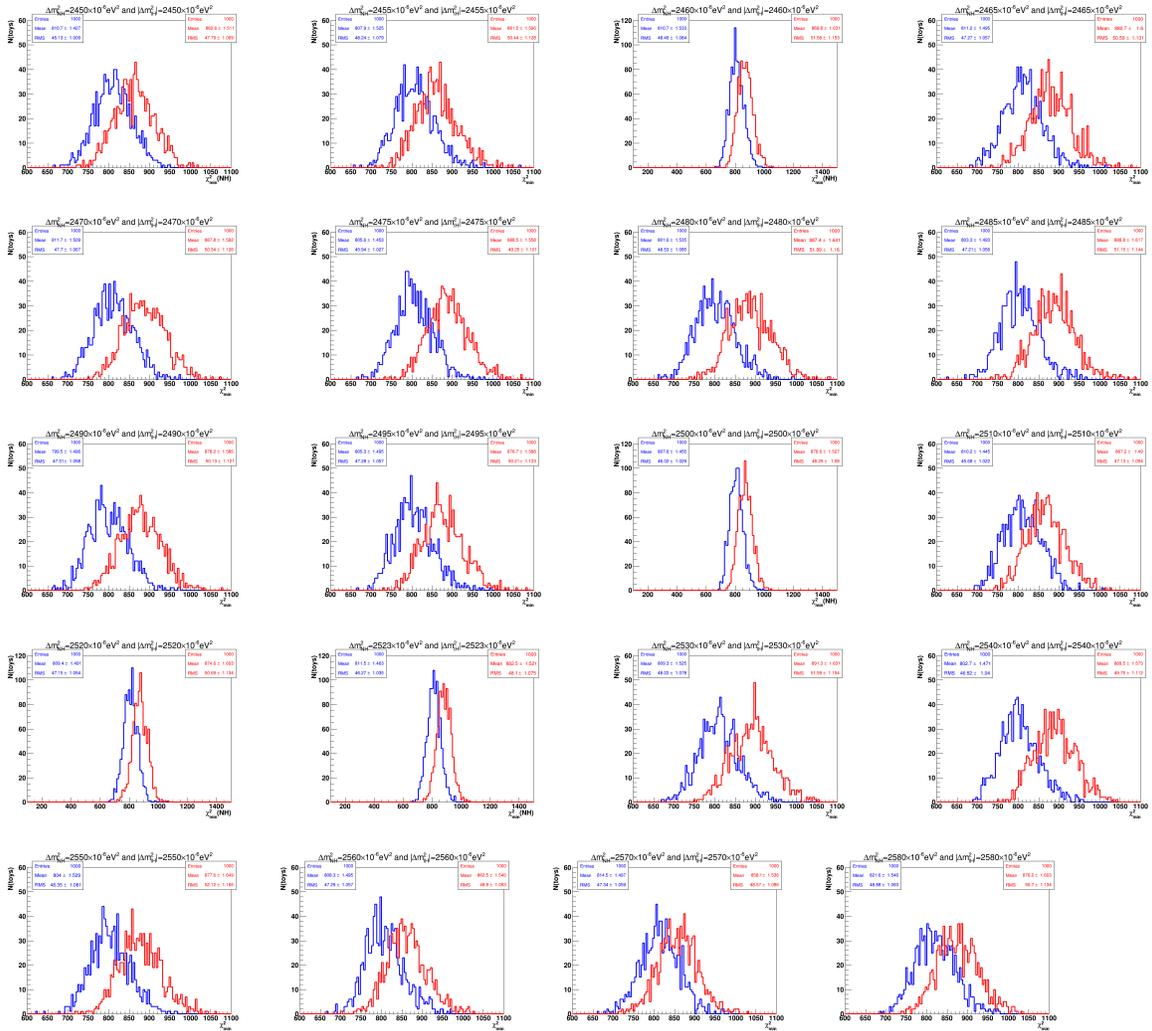

**Figure 21:** $\chi^2$ *distributions for 1000 toy JUNO-like simulations generated for NH samples at 20 different values of the atmospheric mass in the range of* $2.450 \times 10^{-3}\mathrm{eV}^2 \leq |\Delta m^2|_{input} \leq 2.580 \times 10^{-3}\mathrm{eV}^2$ *for NH hypothesis (blue graphs) and for IH hypothesis (red graphs) with six years of exposure and the ten near reactor cores with an infinite energy resolution.*

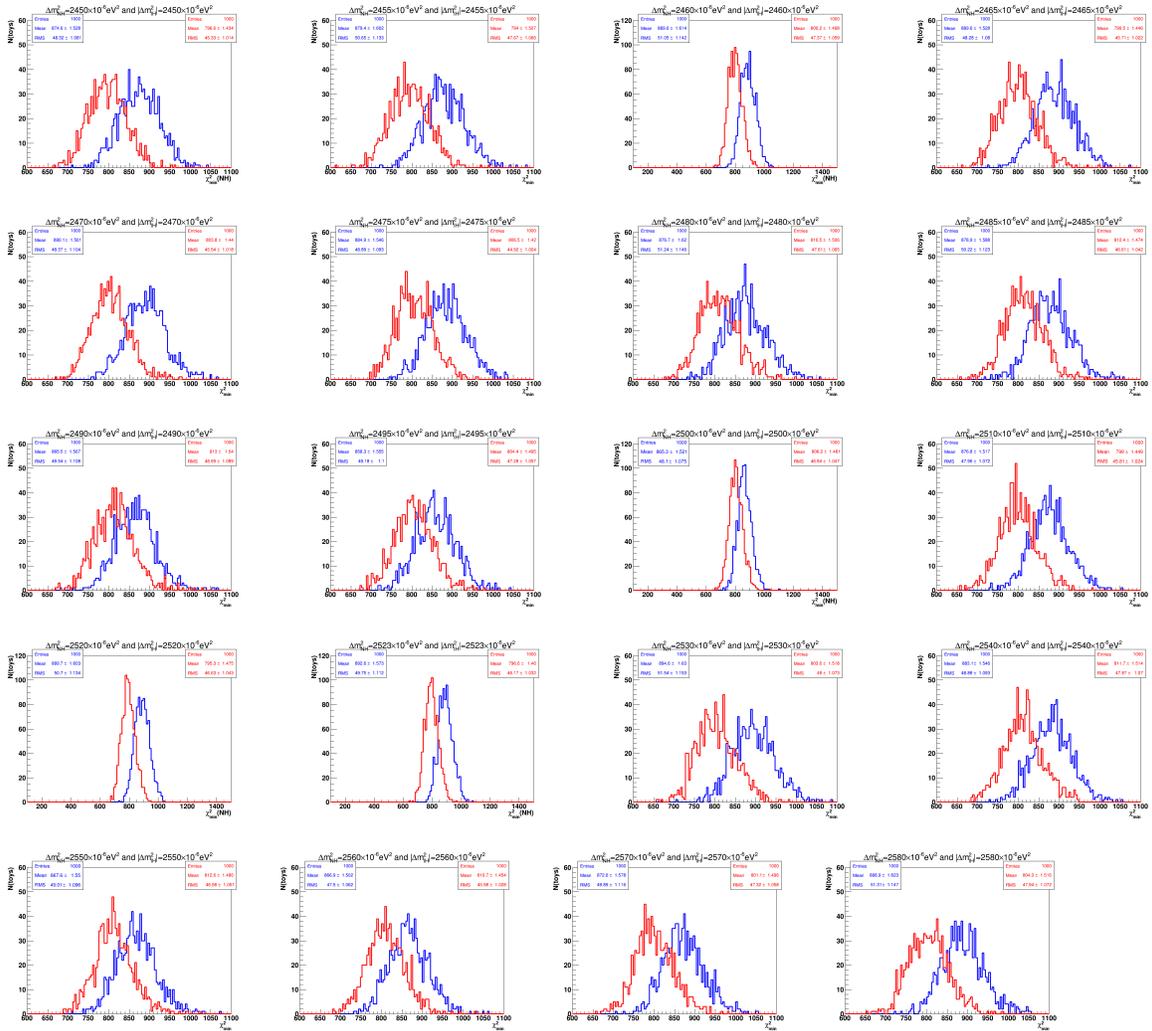

**Figure 22:** $\chi^2$ distributions for 1000 toy JUNO-like simulations generated for IH samples at 20 different values of the atmospheric mass in the range of $2.450 \times 10^{-3}\text{eV}^2 \leq |\Delta m^2|_{input} \leq 2.580 \times 10^{-3}\text{eV}^2$ for NH hypothesis (blue graphs) and for IH hypothesis (red graphs) with six years of exposure and the ten near reactor cores with an infinite energy resolution.

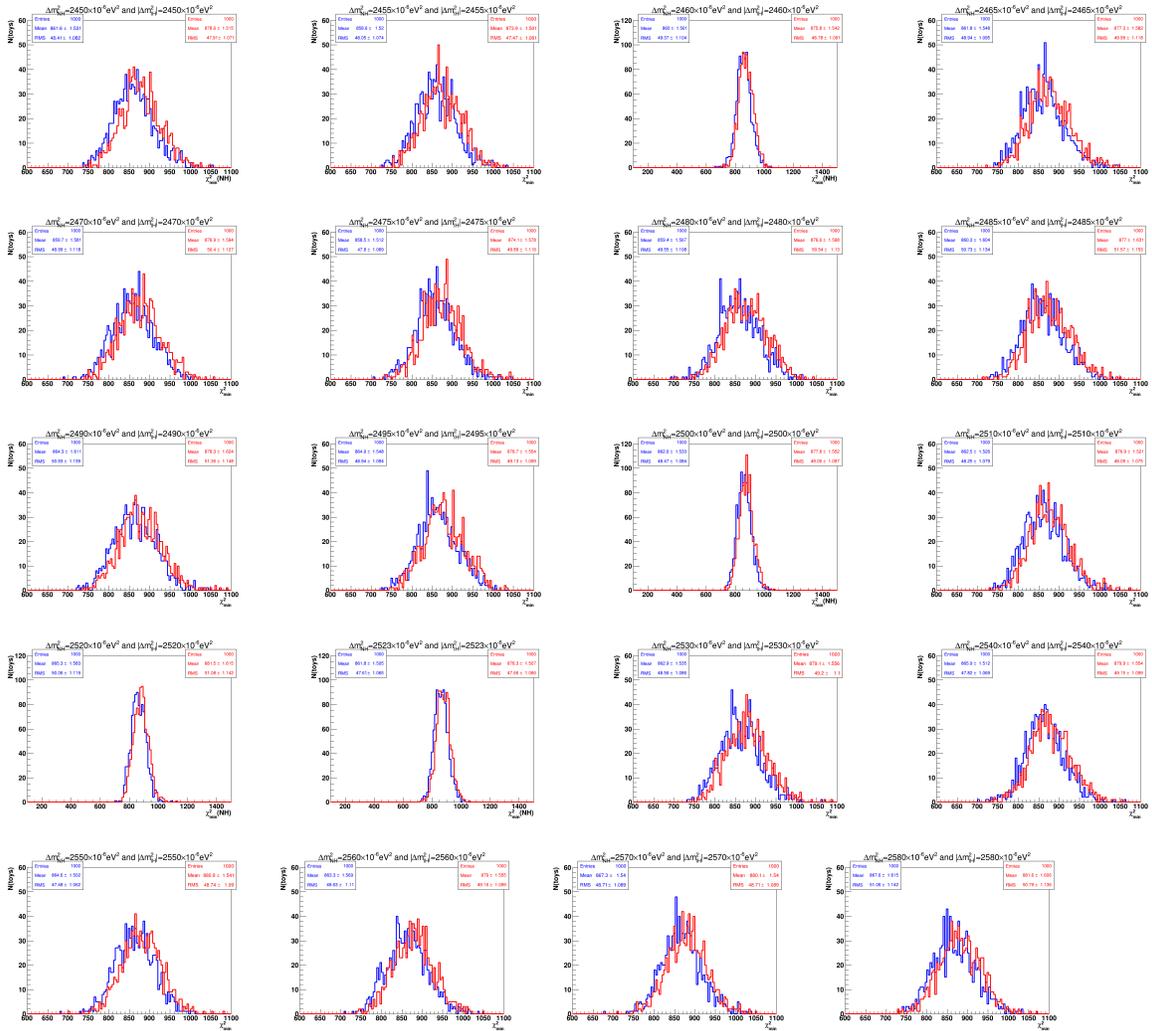

**Figure 23:** $\chi^2$ *distributions for 1000 toy JUNO-like simulations generated for NH samples at 20 different values of the atmospheric mass in the range of* $2.450 \times 10^{-3} \text{eV}^2 \leq |\Delta m^2|_{input} \leq 2.580 \times 10^{-3} \text{eV}^2$ *for NH hypothesis (blue graphs) and for IH hypothesis (red graphs) with six years of exposure and the ten near reactor cores with an* $3\%/\sqrt{E}$ *energy resolution.*

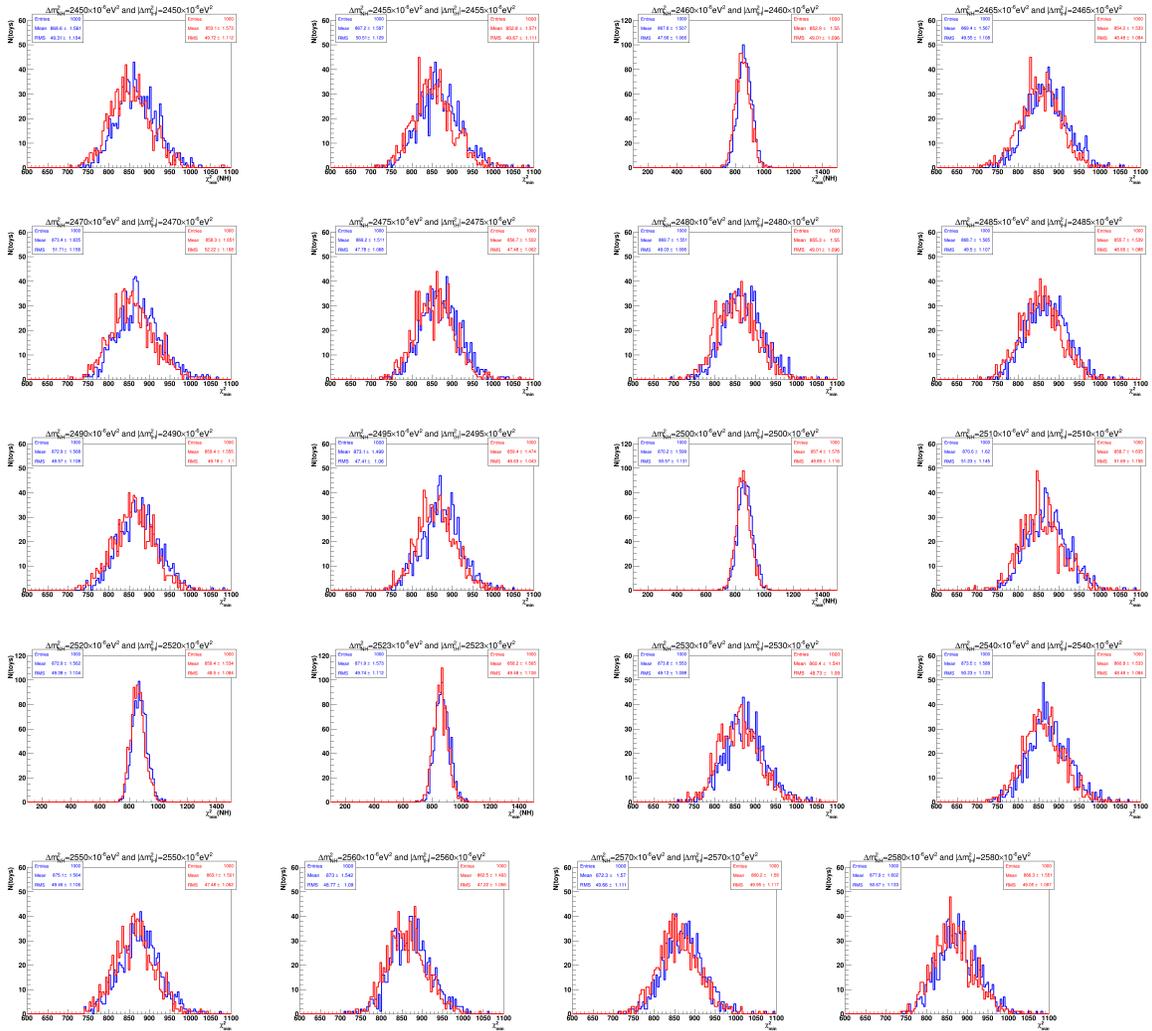

**Figure 24:** $\chi^2$ distributions for 1000 toy JUNO-like simulations generated for IH samples at 20 different values of the atmospheric mass in the range of $2.450 \times 10^{-3}\text{eV}^2 \leq |\Delta m^2|_{input} \leq 2.580 \times 10^{-3}\text{eV}^2$ for NH hypothesis (blue graphs) and for IH hypothesis (red graphs) with six years of exposure and the ten near reactor cores with an $3\%/\sqrt{E}$ energy resolution.